\documentclass[11pt,a4paper]{article}
\usepackage[T1]{fontenc}
\usepackage{microtype}
\usepackage[margin=1in]{geometry}
\usepackage{amsmath,amssymb,amsfonts,mathtools}
\numberwithin{equation}{section}
\usepackage{bm}
\usepackage{array,booktabs,longtable}
\usepackage[natbibapa]{apacite}
\usepackage[hyperfootnotes=false]{hyperref}

\newcommand{\Phie}{{\Phi_e}}
\newcommand{\Gdet}{\sqrt{\lvert\det g\rvert}}

\usepackage{empheq}
\usepackage[most]{tcolorbox}

\usepackage{xcolor}

\usepackage{setspace}
\title{A Five-Dimensional Geometric Formulation of Geophysical Fluid Dynamics}
\author{%
\begin{tabular}{c}
St{\'e}phane Gaudreault$^{1}$\thanks{Corresponding author's email: \href{mailto:stephane.gaudreault@ec.gc.ca}{stephane.gaudreault@ec.gc.ca}}, Seth Taylor$^{2}$, Arad Nasiri$^{3}$,\\
Sanjeev S. Seahra$^{3}$ and Raymond J. Spiteri$^{2}$
\end{tabular}%
\\[1ex]
\begin{minipage}{0.95\textwidth}
\centering\small
$^{1}$Recherche en pr{\'e}vision num{\'e}rique atmosph{\'e}rique, Environnement et Changement climatique Canada, Dorval, Canada\\
$^{2}$Department of Computer Science, University of Saskatchewan, Saskatoon, Canada\\
$^{3}$Department of Mathematics and Statistics, University of New Brunswick, Fredericton, New Brunswick, Canada
\end{minipage}
}
\date{}

\begin{document}
\maketitle

\begin{abstract}
We apply a geometric formulation of Newtonian dynamics to inviscid compressible geophysical flow, with gravity and the effects of rotating or moving coordinates incorporated in the metric tensor. This construction relies on the introduction of an auxiliary fifth coordinate, which allows mass and momentum transport, internal energy, and pressure work to be combined into a single mass--momentum tensor. The covariant divergence of this tensor leads to the continuity equation for mass density, the three momentum equations, and the thermodynamic equation, with the system closed by an equation of state. A time-independent metric also yields conservation of total energy. We recover familiar forms of the equations and illustrate the construction in rotating spherical and terrain-following coordinates. The thin-shell and spherical-geopotential approximations, with constant effective gravity, simplify the metric while retaining vertical gravity and the traditional Coriolis terms. Emphasis is placed on the connection between the geometric structure of the formulation and the equations commonly used in atmospheric dynamics.
\end{abstract}
\vspace{0.2cm}

{\renewcommand{\abstractname}{Résumé}
\begin{abstract}
Nous appliquons à la dynamique des fluides géophysiques une formulation géométrique de la dynamique newtonienne dans laquelle la gravité ainsi que les effets liés à la rotation ou au mouvement du système de coordonnées sont intégrés au tenseur métrique. Cette construction repose sur l’introduction d’une cinquième coordonnée auxiliaire, qui permet de regrouper dans un tenseur de masse--quantité de mouvement le transport de la masse et de la quantité de mouvement, l’énergie interne ainsi que le travail des forces de pression. La divergence covariante de ce tenseur conduit à l’équation de continuité, aux trois équations du mouvement et à l’évolution thermodynamique, le système étant fermé par une équation d’état. Une métrique indépendante du temps conduit également à la conservation de l’énergie totale. Nous retrouvons plusieurs formes usuelles des équations et illustrons la construction en coordonnées sphériques en rotation ainsi qu’en coordonnées verticales suivant le relief. Des approximations de couche mince et de géopotentiel sphérique, avec une gravité effective constante, simplifient la métrique tout en conservant la gravité verticale et les termes de Coriolis traditionnels. L’objectif est de mettre en évidence le lien entre la structure géométrique de la formulation et les équations couramment utilisées en dynamique atmosphérique.
\end{abstract}
}

\par

\section{Introduction}
\label{sec:introduction}

Writing down the equations governing atmospheric motion is one thing; writing them down in a coordinate system fit for computation is quite another. The atmosphere is a compressible fluid, rotating with the planet and stratified by gravity. Operational numerical weather prediction commonly employs spherical coordinates rotating with the Earth, often combined with a terrain-following vertical coordinate. These modelling choices introduce many coordinate- and frame-dependent terms. Model developers frequently need to change coordinate systems or simplify the equations while preserving consistency with Newtonian mechanics. Deriving or modifying the individual terms separately is error-prone and can compromise conservation properties. Here, we present a geometric formulation that represents these terms, including Coriolis and centrifugal accelerations, through the metric tensor and its associated connection. This formulation provides a set of rules for transforming the equations and treating the geometric terms consistently.

These transformation rules are provided by tensor analysis \citep{Charron2014}. Tensor equations are \emph{covariant}: their mathematical form is preserved under a change of coordinates or reference frame. The tensor transformation rules ensure that these equations represent the same physical statements in different coordinate systems. Once the atmospheric equations are written in tensor form, the metric and inertial terms associated with a change of coordinates or reference frame can be derived systematically. Several four-dimensional tensor formulations of the fluid equations have been developed along these lines \citep{Vinokur1974,Avis1976,Charron2014,pudykiewicz2017metric}. In particular, \citet{Charron2014} showed how the governing equations used in atmospheric models can be obtained from a compact four-dimensional tensor formulation.

In the four-dimensional formulation of \citet{Charron2014}, the coordinate and inertial terms are treated systematically, but gravity remains an explicit external force. We extend this formulation by incorporating the gravitational force into the geometry, while retaining the ability to use rotating and terrain-following coordinates alongside their approximations. A geometric description of Newtonian gravity for point particles was introduced by {\'E}lie Cartan, who represented the trajectories of freely falling particles as geodesics of a four-dimensional connection \citep{cartan1923varietes,cartan1924varietes}. This formalism incorporates Newtonian gravity into the geometry, but the gravitational connection is not determined by a single non-degenerate four-dimensional metric tensor. Introducing a fifth dimension allows such a metric description, as shown by \citet{Eisenhart1928}. In this construction, Newtonian trajectories in a gravitational potential are realized as the projection of geodesics in the extended five-dimensional geometry to the four-dimensional Newtonian space-time. The auxiliary coordinate, denoted by $x^4$, introduces an additional degree of freedom allowing gravity to be incorporated into the five-dimensional metric structure. It does not correspond to a direction in physical space or time. Translation invariance in this coordinate introduces a symmetry associated with the conservation of mass. A modern differential-geometric treatment of this idea was later developed by \citet{Duval1985,Duval1991}.

The Eisenhart--Duval treatment of gravity can be applied to fluid dynamics by introducing a symmetric mass--momentum tensor $T^{AB}$ that represents mass and momentum transport as well as internal energy and pressure work \citep{banerjee2016null}. On the extended five-dimensional manifold, the conservation laws can be expressed compactly through its covariant divergence,
\begin{equation}
\nabla_A T^{AB}=0.
\end{equation}
This expression gives the continuity equation, the three momentum equations, and the thermodynamic equation. Together with an equation of state, these equations form a closed system of partial differential equations.

The aim of this paper is to present an explicit five-dimensional formulation of compressible atmospheric dynamics in rotating and moving coordinates, building on the Eisenhart--Duval construction and the null-fluid formulation \citep{Eisenhart1928,Duval1985,Duval1991,banerjee2016null}. The main contribution is the derivation of the metric, geometric source terms, and balance laws for the coordinate systems and approximations used in atmospheric modelling. We construct the rotating spherical metric and its transformation to a generalized, possibly time-dependent vertical coordinate, and derive the corresponding continuity, momentum, internal-energy, and total-energy balances. We also show how a thin-shell metric, combined with a spherical effective potential and constant effective gravity, recovers the nonhydrostatic shallow-atmosphere equations with traditional Coriolis terms. These results extend the atmospheric tensor formulation of \citet{Charron2014} by incorporating gravity into the metric and treating its consequences for the governing equations and their approximations.

The paper is organized as follows. Sections~\ref{sec:manifold}--\ref{sec:synchronous} introduce the geometric formulation and the kinematic relations. Sections~\ref{sec:tensor} and~\ref{sec:forms} present the fluid equations and their alternative forms. Sections~\ref{sec:sphere} and~\ref{sec:vertical} discuss applications to rotating spherical and generalized vertical coordinates. Section~\ref{sec:thinshell} introduces the thin-shell and spherical-geopotential approximations, with constant effective gravity, and derives the resulting equations with traditional Coriolis terms. The integral form of the balance laws is given in Appendix~\ref{sec:gauss}, and a list of symbols is provided in Appendix~\ref{sec:symbols}.

\section{Geometric Preliminaries}
\label{sec:manifold}

We introduce the tensor calculus and geometric identities needed to formulate the fluid equations. Familiarity with multivariable calculus is assumed but no prior knowledge of differential geometry is required. For a more comprehensive mathematical treatment, the reader is referred to \citet{isham1989modern}.

\subsection{Manifolds, coordinate systems, and tensors}

An $n$-dimensional \emph{smooth manifold} $\mathcal{M}$ is a topological space that locally resembles Euclidean space $\mathbb{R}^n$, meaning that every point has a neighbourhood that maps bijectively and continuously onto an open subset of $\mathbb{R}^n$ with a continuous inverse. Such a mapping is called a \emph{chart}. A collection of compatible charts covering the manifold is called an \emph{atlas}. Each chart assigns coordinates to every point in its domain. Some charts in the atlas overlap on their common domains, and on each overlap, the two sets of coordinates are related by smooth invertible functions called \emph{transition maps}.

Throughout this paper, we adopt the following index conventions: uppercase Latin indices $A,B,C,\ldots$ take values in $\{0, 1, 2, 3, 4\}$; lowercase Greek indices $\mu,\nu,\ldots$ take values in $\{0, 1, 2, 3\}$; and lowercase Latin indices $i,j,k,\ldots$ take values in $\{1, 2, 3\}$. An index appearing once as a superscript and once as a subscript is summed over the range appropriate to that index.

In the five-dimensional setting used in this study, a general transformation from a coordinate system $S$, with coordinates $x^0, x^1, x^2, x^3, x^4$, to $\tilde{S}$, with coordinates $\tilde{x}^0, \tilde{x}^1, \tilde{x}^2, \tilde{x}^3, \tilde{x}^4$, takes the form
\begin{align}
  \tilde x^A &= \tilde x^A(x^0, x^1, x^2, x^3, x^4),
  \label{eq:gen_transform}
\end{align}
The transformation is required to be smooth and invertible on the overlap of the coordinate charts. Further restrictions can be imposed depending on the physical properties of the system. In Section~\ref{sec:newtonian_geometry}, we restrict these transformations to preserve the absolute time of Newtonian mechanics and leave the auxiliary coordinate unchanged.

Physical quantities are described by fields that assign a scalar, vector, or tensor to each point of the manifold. Their components transform according to definite rules under a change of coordinates, so that the physical quantities they represent remain unchanged.

A \emph{scalar} is a quantity whose numerical value at a point is the same in all coordinate systems. Pressure, temperature, and density are examples of scalars. A \emph{vector} at a point $P$ is defined as a tangent to a curve passing through $P$. If we denote its components in the coordinate-induced basis as $V^A$, they transform contravariantly between two coordinate systems:
\begin{equation}
\tilde V^B = \frac{\partial\tilde x^B}{\partial x^A}\,V^A.
\label{eq:contravariant_transf}
\end{equation}
A complementary object is the \emph{covector} (or covariant vector), of which the gradient of a scalar field is an example. For a scalar field $f$, the components $W_A = \partial f/\partial x^A$ transform according to
\begin{equation}
\tilde W_B = \frac{\partial x^A}{\partial\tilde x^B}\,W_A.
\label{eq:covariant_transf}
\end{equation}

Higher-rank tensors are defined by requiring their components to transform as products of the Jacobian matrix $\partial\tilde x^A/\partial x^B$ (for each upper index) and its inverse $\partial x^A/\partial\tilde x^B$ (for each lower index). For example, a rank-two tensor with one contravariant and one covariant index transforms as
\begin{equation*}
    \widetilde{Q}^{A}{}_{B}
    =
    \frac{\partial \widetilde{x}^{A}}{\partial x^{C}}
    \frac{\partial x^{D}}{\partial \widetilde{x}^{B}}
    Q^{C}{}_{D}.
\end{equation*}
Each upper index contributes a factor of the Jacobian, whereas each lower index contributes a factor of the inverse Jacobian.

A quantity is a tensor if and only if its components transform according to this rule.\footnote{This follows the classical component-based approach often termed the ``physicist's definition'' of a tensor. In modern differential geometry, a tensor is intrinsically defined as a multilinear map independent of any choice of coordinates. The two definitions are equivalent: the transformation rule simply describes how the components of such a multilinear map change when expressed in a coordinate-induced basis.}

\subsection{Metric tensor}

A central geometric object in what follows is the \emph{metric tensor} $g_{AB}$. It can be understood through the \emph{line element}, which assigns a squared interval to an infinitesimal coordinate displacement:
\begin{equation*}
  \mathrm d\ell^2=g_{AB}\,\mathrm dx^A\,\mathrm dx^B.
\end{equation*}
In Euclidean space expressed in Cartesian coordinates, the spatial line element is the squared distance $\mathrm d\ell_{\mathrm{space}}^2=\mathrm dx^2+\mathrm dy^2+\mathrm dz^2$. In spherical coordinates with longitude $\lambda$, latitude $\varphi$, and radius $r$, it becomes
\begin{equation*}
  \mathrm d\ell_{\mathrm{space}}^2
  =r^2\cos^2\varphi\,\mathrm d\lambda^2
  +r^2\,\mathrm d\varphi^2+\mathrm dr^2.
\end{equation*}
The metric coefficients convert coordinate increments into physical distances. \par

The metric defines a smoothly varying inner product on the tangent space of the manifold, assigning a scalar $g_{AB}V^AW^B$ to any pair of tangent vectors $V^A$ and $W^A$ defined at the same point. This provides a correspondence between vectors and covectors: contracting a vector with the metric lowers an index, $V_A=g_{AB}V^B$, whereas contraction with the inverse metric $g^{AB}$, defined by $g^{AC}g_{CB}=\delta^A{}_B$, raises indices according to $V^A=g^{AB}V_B$.

In the Eisenhart--Duval construction, the metric is a \emph{pseudo-Riemannian metric}, i.e. it is symmetric and non-degenerate but need not be positive definite. Its signature specifies the numbers of positive and negative directions in a tangent-space basis that diagonalizes the metric. Unlike the Euclidean dot product, however, a pseudo-Riemannian metric need not assign a positive squared norm to every non-zero vector.\par

The metric can be used to define a volume associated with a coordinate cell. Whereas the product of coordinate increments gives the size of a cell in coordinate space, the metric supplies the factor that converts this coordinate volume into geometric volume. In five dimensions, the resulting volume element is
\begin{equation*}
  \mathrm dV_5=\Gdet\,\mathrm d^5x,
\end{equation*}
where $\det g$ is the determinant of the covariant metric tensor in the chosen coordinates and $\mathrm d^5x$ is the coordinate volume measure. The absolute value accounts for the metric's indefinite signature. This volume element is invariant under a coordinate transformation: although the coordinate cell and the metric factor change, their product represents the same geometric volume.

\subsection{Affine connection} \label{sec:connection}

An \emph{affine connection} provides a rule for comparing vectors at neighbouring points of the manifold and for differentiating tensor fields in a geometrically meaningful way.

Ordinary partial derivatives of tensor components do not in general yield tensors because the coordinate basis vectors themselves vary from point to point. The affine connection defines the corresponding \emph{covariant derivative}, which accounts for this variation. For a vector $V^A$, the covariant derivative in the $B$-direction is
\begin{equation}
  \nabla_B V^A = \frac{\partial V^A}{\partial x^B} + \Gamma^A_{BC}\,V^C,
  \label{eq:cov_deriv}
\end{equation}
where $\nabla_B$ denotes covariant differentiation, whereas $\partial V^A/\partial x^B$ is the ordinary partial derivative. Similarly, the covariant derivative of a rank-two contravariant tensor $T^{AB}$ in the $C$-direction is
\begin{equation}
  \nabla_C T^{AB} = \frac{\partial T^{AB}}{\partial x^C} + \Gamma^A_{CD}T^{DB} + \Gamma^B_{CD}T^{AD}
  \label{eq:cov_deriv_tensor}
\end{equation}
Note that neither the partial derivatives of the components nor the Christoffel symbols in \eqref{eq:cov_deriv} and \eqref{eq:cov_deriv_tensor} transform tensorially; only the complete sum of terms in each expression obeys the tensor transformation law. \par

The metric tensor $g_{AB}$ determines a unique torsion-free, metric-compatible affine connection, called the \emph{Levi--Civita connection}. Metric compatibility means that the covariant derivative of the metric vanishes:
\begin{equation*}
  \nabla_C g_{AB}=0.
\end{equation*}
Geometrically, this means that parallel transport preserves the scalar product of vectors. Algebraically, raising or lowering indices before or after covariant differentiation yields the same result. The coefficients of the Levi--Civita connection in a coordinate basis are the \emph{Christoffel symbols of the second kind}:
\begin{equation}
  \Gamma^A_{BC} = \frac{1}{2}g^{AD}\!\left(\frac{\partial g_{DB}}{\partial x^C} + \frac{\partial g_{DC}}{\partial x^B} - \frac{\partial g_{BC}}{\partial x^D}\right).
  \label{eq:christoffel}
\end{equation}
The torsion-free condition appears as symmetry in the two lower indices, $\Gamma^A_{BC}=\Gamma^A_{CB}$. \par

The covariant divergence of a vector is obtained by contracting the indices of the covariant derivative \eqref{eq:cov_deriv}. Using the identity
\begin{equation*}
\frac{1}{\Gdet}\frac{\partial\Gdet}{\partial x^A} = \Gamma^C_{AC},    
\end{equation*}
this expression becomes
\begin{equation*}
    \nabla_A V^A = \frac{1}{\Gdet}\frac{\partial}{\partial x^A}\bigl(\Gdet\,V^A\bigr).
\end{equation*}
Similarly, contracting the derivative index with a tensor index in \eqref{eq:cov_deriv_tensor} defines the covariant divergence
\begin{equation}
  \nabla_A T^{AB}
  = \frac{1}{\Gdet}\,\frac{\partial}{\partial x^A}\bigl(\Gdet\,T^{AB}\bigr) + \Gamma^B_{AC}\,T^{AC}\,,
  \label{eq:divT}
\end{equation}
which is used below to express local balance laws in terms of mass and momentum transport.

\section{Five-Dimensional Newtonian Geometry}
\label{sec:newtonian_geometry}

In Newtonian mechanics, a particle subject to no force moves with constant velocity along a straight line in an inertial frame. Gravity changes this motion through a force derived from a potential. The Eisenhart--Duval construction incorporates this potential into a five-dimensional metric, so that a particle moving under gravity alone follows a geodesic of the extended geometry. This gives a geometric description of gravitational motion while retaining Newtonian mechanics. Time remains absolute and spatial distances at fixed time retain their Euclidean meaning.

\subsection{The Eisenhart--Duval metric}

We use coordinates $x^A=(x^0,x^1,x^2,x^3,x^4)$, where $x^0=t$ and $x^4$ denotes the auxiliary coordinate. We restrict coordinate changes to \emph{synchronous transformations}, which preserve absolute time and leave the auxiliary coordinate unchanged:
\begin{equation}
  \tilde x^0 = x^0, \qquad
  \tilde x^i = \tilde x^i(x^0,x^1,x^2,x^3), \qquad
  \tilde x^4 = x^4.
  \label{eq:synchronous}
\end{equation}
Notice that the spatial coordinates may depend on time to allow for rotating frames and moving coordinates.

Let $\Phi$ be the Newtonian gravitational potential, a scalar under these transformations. In an admissible coordinate system, the line element takes the form
\begin{equation}
  \mathrm{d}\ell^2 = g_{AB}\,\mathrm{d}x^A\,\mathrm{d}x^B
  = g_{ij}\,\mathrm{d}x^i\,\mathrm{d}x^j
  + 2g_{0i}\,\mathrm{d}x^i\,\mathrm{d}x^0
  + \left(g^{ij}g_{0i}g_{0j}-2\Phi\right)(\mathrm{d}x^0)^2
  + 2\,\mathrm{d}x^0\,\mathrm{d}x^4,
  \label{eq:bargmann_metric}
\end{equation}
where the spatial block $g_{ij}$ is a positive-definite $3$-metric, $g^{ij}$ is its inverse, and the time--space components $g_{0i}$ describe frame motion, such as rotation. The term $g^{ij}g_{0i}g_{0j}$ represents the frame-motion contribution to the time coefficient. At fixed time, the line element reduces to $g_{ij}\,\mathrm dx^i\,\mathrm dx^j$, so the spatial block measures physical distances independently of the auxiliary coordinate.

We define the effective potential by
\begin{equation}
  \Phie\equiv\Phi-\frac{1}{2}g^{ij}g_{0i}g_{0j},
  \qquad g_{00}=-2\Phie.
  \label{eq:Phie_split}
\end{equation}
Unlike the Newtonian gravitational potential $\Phi$, the effective potential $\Phie$ is not generally a scalar under synchronous transformations, because it also includes a contribution from frame motion. For instance, in a stationary rotating frame, the additional contribution is the centrifugal potential. 

The components of the covariant metric tensor are
\begin{equation}
  g_{00} = -2\Phie, \quad
  g_{0i} = g_{i0}, \quad
  g_{04} = g_{40} = 1, \quad
  g_{i4} = g_{4i} = 0, \quad
  g_{44} = 0,
  \label{eq:g_covariant}
\end{equation}
which, in block form with rows and columns ordered $(0, 1, 2, 3, 4)$, read
\begin{equation}
  g_{AB} =
  \begin{pmatrix}
    -2\Phie & g_{01} & g_{02} & g_{03} & 1 \\
    g_{01}    & g_{11} & g_{12} & g_{13} & 0 \\
    g_{02}    & g_{12} & g_{22} & g_{23} & 0 \\
    g_{03}    & g_{13} & g_{23} & g_{33} & 0 \\
    1      & 0 & 0 & 0 & 0
  \end{pmatrix}.
  \label{eq:g_block}
\end{equation}
Since the spatial metric $g_{ij}$ is positive definite, the five-dimensional metric has signature $(-,+,+,+,+)$. The squared interval can be positive, negative, or zero, and a non-zero vector can have zero norm. Such a vector is called \emph{null}; its existence does not imply that the metric is degenerate. The null condition will be used below to relate the fifth velocity component to the mechanical Lagrangian.

The auxiliary coordinate $x^4$ labels points along a \emph{fibre} attached to each physical spacetime event. Points differing only in $x^4$ correspond to the same physical position and time. Thus, we take the metric coefficients and physical fields to be independent of $x^4$, although this coordinate may vary along a trajectory. A displacement along this fibre has zero squared interval because $g_{44}=0$, so the auxiliary direction is null. The full metric remains non-degenerate, however, through the coupling $g_{04}=1$ between time and the auxiliary coordinate. 

\subsection{Contravariant metric}

The contravariant metric $g^{AB}$ is defined as the inverse of the covariant metric tensor by $g^{AC}g_{CB}=\delta^A{}_B$. Its components can be obtained directly from the structure of the fifth row and column of $g_{AB}$. \par

The fifth column of the covariant metric has only one non-zero entry, $g_{04}=1$. Setting $B=4$ in the inverse relation gives
\begin{equation*}
  g^{AC}g_{C4}=g^{A0}=\delta^A{}_4.
\end{equation*}
Thus, using the symmetry of the inverse metric, we obtain its time components:
\begin{equation}
  g^{00}=0,\qquad g^{0i}=0,\qquad g^{04}=g^{40}=1.
  \label{eq:g_inv_t}
\end{equation}

For the spatial components, the inverse relation gives
\begin{equation*}
  g^{i0}g_{0j}+g^{ik}g_{kj}+g^{i4}g_{4j}
  =\delta^i{}_j.
\end{equation*}
Since $g^{i0}=0$ and $g_{4j}=0$, this reduces to
\begin{equation*}
  g^{ik}g_{kj}=\delta^i{}_j.
\end{equation*}
The spatial block of the inverse metric is therefore the inverse of $g_{ij}$. 

To obtain the remaining components, we use the equivalent identity with the multiplication order reversed, $g_{AC}g^{CB}=\delta_A{}^B$. In particular, $g_{0A}g^{Ai}=0$ gives $g_{0k}g^{ki}+g^{4i}=0$; hence
\begin{equation}
  g^{4i} = g^{i4} = -g^{ij}g_{0j}.
  \label{eq:g_inv_5i}
\end{equation}
Finally, $g_{0A}g^{A4}=0$ gives $-2\Phie+g_{0i}g^{i4}+g^{44}=0$. Substituting \eqref{eq:g_inv_5i} yields $g^{44}=g_{0i}g^{ij}g_{0j}+2\Phie$. In block form, the inverse metric is given by
\begin{equation}
  g^{AB} =
  \begin{pmatrix}
    0 & 0 & 0 & 0 & 1 \\
    0 & g^{11} & g^{12} & g^{13} & -g^{1m}g_{0m} \\
    0 & g^{12} & g^{22} & g^{23} & -g^{2m}g_{0m} \\
    0 & g^{13} & g^{23} & g^{33} & -g^{3m}g_{0m} \\
    1 & -g^{1m}g_{0m} & -g^{2m}g_{0m} & -g^{3m}g_{0m} & g_{0i}g^{ij}g_{0j} + 2\Phie
  \end{pmatrix}.
  \label{eq:g_inv_block}
\end{equation}
Substituting the definition of $\Phie$ from \eqref{eq:Phie_split} gives $g^{44}=2\Phi$. Thus, $g^{44}$ transforms as a scalar under the synchronous transformations considered here.

\subsection{Example: particle in a gravitational potential}
\label{subsec:particle}

The sign of the gravitational term is fixed by the requirement that free-fall trajectories be geodesics. We verify this by recovering Newtonian motion for a particle in a gravitational potential. Consider a particle moving in a static gravitational potential $\Phie=\Phi$ in an inertial frame ($g_{0i}=0$) and use Cartesian spatial coordinates. In Cartesian coordinates, the Euclidean spatial metric has components $\delta_{ij}=\operatorname{diag}(1,1,1)$, with inverse $\delta^{ij}$. The five-dimensional metric then reduces to \begin{equation*}
  \mathrm{d}\ell^2
  = \delta_{ij}\,\mathrm{d}x^i\mathrm{d}x^j
  -2\Phi\,\mathrm{d}t^2
  +2\,\mathrm{d}t\,\mathrm{d}x^4.
\end{equation*}

Let $x^A(t)$ denote the coordinates of the lifted particle trajectory and write $u^A=\mathrm dx^A/\mathrm dt$, so $u^0=1$. For this metric, $\Gamma^0_{AB}=0$, so the time coordinate itself can serve as an affine parameter along every geodesic on which it is not constant, giving the geodesic equation
\begin{equation}
  \frac{\mathrm d^2x^A}{\mathrm dt^2}
  +\Gamma^A_{BC}\frac{\mathrm dx^B}{\mathrm dt}\frac{\mathrm dx^C}{\mathrm dt}=0.
  \label{eq:geodesic}
\end{equation}
In this Cartesian, non-rotating case, the spatial inverse components are $g^{ij}=\delta^{ij}$ and $g^{i0}=g^{i4}=0$. Substitution into \eqref{eq:christoffel} gives the only non-zero spatial acceleration coefficient:
\begin{equation*}
  \Gamma^i_{00}
  = -\frac{1}{2}\delta^{ij}\frac{\partial g_{00}}{\partial x^j}
  = \delta^{ij}\frac{\partial\Phi}{\partial x^j}.
\end{equation*}
Since $u^0=1$, the spatial geodesic equation becomes
\begin{equation*}
  \frac{\mathrm{d}^2x^i}{\mathrm{d}t^2}
  = -\Gamma^i_{00}
  = -\delta^{ij}\frac{\partial\Phi}{\partial x^j},
\end{equation*}
which recovers Newton's second law for a particle in a gravitational potential, with the gravitational acceleration represented by the metric connection. The particle acceleration along the fibre is not zero, however, because $\Gamma^4_{0i}=\Gamma^4_{i0}=-\frac{\partial\Phi}{\partial x^i}$. Hence, we have
\begin{equation}
\label{eq:d2x4}
    \frac{d^2x^4}{dt^2}=2u^i\frac{\partial\Phi}{\partial x^i}.
\end{equation}
The auxiliary velocity can be chosen so that the lifted trajectory has zero squared norm. With $K=\tfrac12\delta_{ij}u^iu^j$, this condition reads $2K-2\Phi+2u^4=0$, giving $u^4=\Phi-K$. Newton's equation gives $\mathrm dK/\mathrm dt=-\mathrm d\Phi/\mathrm dt$, so
\begin{equation*}
  \frac{\mathrm d^2x^4}{\mathrm dt^2}
  =\frac{\mathrm d}{\mathrm dt}(\Phi-K)
  =2\frac{\mathrm d\Phi}{\mathrm dt}
  =2u^i\frac{\partial\Phi}{\partial x^i}.
\end{equation*}
This agrees with the fifth geodesic equation \eqref{eq:d2x4}. Thus, for a freely falling particle, the auxiliary evolution specified by the null condition is consistent with geodesic motion and imposes no additional restriction on the Newtonian trajectory.

\section{Kinematics and Spatial Geometry}
\label{sec:synchronous}

\subsection{The five-velocity and its mechanical interpretation}

A parcel trajectory is a curve whose coordinates are $x^A(t)$. Its tangent vector, the \emph{five-velocity}, has components
\begin{equation}
  u^A = \frac{\mathrm{d}x^A}{\mathrm{d}t}.
  \label{eq:5vel}
\end{equation}

Since the trajectory is parametrized by time, the zeroth component of the five-velocity is $u^0=\mathrm dt/\mathrm dt=1$. Consequently, differentiation along the trajectory includes both local time evolution and spatial advection. For a scalar field $f$ independent of $x^4$, the material derivative along the fluid trajectory is given by
\begin{equation*}
  \frac{df}{dt}\equiv u^A\frac{\partial f}{\partial x^A}=\frac{\partial f}{\partial t}+u^i\frac{\partial f}{\partial x^i}.
\end{equation*}
The three components $u^i=\mathrm dx^i/\mathrm dt$ are coordinate velocities of the parcel.

The fifth component, $u^4=\mathrm dx^4/\mathrm dt$, describes motion along the auxiliary coordinate, whose evolution must be specified. Following the null-geodesic formulation of \citet{Duval1991}, we choose this evolution so that the five-velocity is null:
\begin{equation}
  g_{AB}u^Au^B = 0.
  \label{eq:null_norm}
\end{equation}
Once the spatial velocity and the metric are known, this condition determines $u^4$. A physical trajectory is extended to the five-dimensional manifold by using the null condition to fix the evolution of the auxiliary coordinate, which introduces no restrictions on the physical motion itself.

To interpret this condition, consider the four-dimensional block of the metric. With $u^0=1$ and the metric components in \eqref{eq:g_covariant}, we define
\begin{equation}
  L
  \equiv \frac{1}{2}g_{\mu\nu}u^\mu u^\nu
  = \frac{1}{2}g_{ij}u^iu^j
  + g_{0i} u^i
  - \Phie,
  \label{eq:Lagrangian}
\end{equation}
which is the Lagrangian per unit mass in these coordinates. Given that $g_{i4}=g_{44}=0$ and $g_{04}=1$, the null condition \eqref{eq:null_norm} reduces to $g_{\mu\nu}u^\mu u^\nu+2u^4=0$, and therefore
\begin{equation}
  u^4
  = -L
  = \Phie
  - g_{0i} u^i
  - \frac{1}{2}g_{ij}u^iu^j.
  \label{eq:u5}
\end{equation}
The fifth component represents the negative Lagrangian per unit mass rather than an additional physical speed. Because synchronous transformations leave $x^4$ unchanged, $g_{\mu\nu}$ transforms as a four-dimensional tensor under them, so $L$ and $u^4$ are both invariant.

Lowering the time index gives a second useful quantity:
\begin{equation}
  u_0=g_{0A}u^A
  =-\frac12g_{ij}u^iu^j-\Phie
  =-E_{\mathrm{mech}},
  \qquad E_{\mathrm{mech}}=\frac12g_{ij}u^iu^j+\Phie.
  \label{eq:Emech}
\end{equation}
Here $E_{\mathrm{mech}}$ is the mechanical energy per unit mass associated with the chosen coordinates. In rotating coordinates, it contains the kinetic energy relative to the rotating frame and the effective potential, including the centrifugal contribution.

\subsection{Intrinsic derivative}

The covariant derivative along a fluid trajectory defines the \emph{intrinsic derivative}. For any five-vector $V^A$, it is given by
\begin{equation}
  \frac{\mathrm{D}V^A}{\mathrm{D}t}
  := u^B \nabla_B V^A
  = \frac{\mathrm{d}V^A}{\mathrm{d}t}
  + \Gamma^A_{BC}u^B V^C,
\end{equation}
where $\mathrm{d}/\mathrm{d}t = u^B\partial/\partial x^B$ denotes the ordinary derivative along the trajectory. For a scalar field $f$ there are no connection terms, so the two derivatives coincide, $\mathrm Df/\mathrm Dt=\mathrm df/\mathrm dt$. We write $\mathrm D/\mathrm Dt$ for tensors and scalars and reserve $\mathrm d/\mathrm dt$ for the ordinary derivative of individual components and coordinates along the trajectory, which in general is not a tensor.

The independence of the geometry from $x^4$, together with the constant components $g_{4A}=\delta^0_A$, implies that all Christoffel symbols with a lower $4$ index vanish:
\begin{equation}
\Gamma^A_{B4} = \Gamma^A_{4B} = 0 \quad\text{for all }A,B.
\label{eq:Gamma_lower5}
\end{equation}

For a vector field independent of $x^4$, this result gives $\nabla_4 V^A=0$. The intrinsic derivative simplifies to $\mathrm DV^A/\mathrm Dt=u^\mu\nabla_\mu V^A$, with the contraction restricted to the four spacetime indices $\mu=0,1,2,3$.

\section{Fluid dynamics}
\label{sec:tensor}

\subsection{The mass--momentum tensor}

The conservation laws of fluid dynamics describe how mass, momentum, and energy are transported through space and evolve in time. In the five-dimensional formulation, we assemble their densities and fluxes into a symmetric mass--momentum tensor $T^{AB}$. It must account for mass and momentum transport, internal energy, and work done by pressure forces.

Consider an inviscid fluid with density $\rho$, specific internal energy $e$, pressure $p=p(\rho,e)$ supplied by an equation of state, and velocity field $u^A$.

We arrange these quantities in a tensor $T^{AB}$. In the balance equation for a fixed value of the free index $B$, $T^{0B}$ represents the local density and $T^{jB}$ its spatial flux. The components needed for mass and momentum are
\begin{equation*}
  T^{00}=\rho,\qquad
  T^{j0}=\rho u^j,\qquad
  T^{0i}=\rho u^i,\qquad
  T^{ji}=\rho u^ju^i+pg^{ji}.
\end{equation*}
Here, $\rho u^i$ represents both the mass flux and the momentum density. The momentum flux $T^{ji}$ combines transport by fluid motion, $\rho u^ju^i$, with the isotropic pressure stress, $pg^{ji}$.

These components describe mass and momentum transport, but do not yet account for energy transport. The fifth row and column provide the additional components needed for internal energy and pressure work. The kinematic component $u^4=-L$ remains fixed by the null condition, whereas $e$ is an independent thermodynamic variable. The transport term $\rho u^Au^B$ must therefore be supplemented by an internal-energy contribution.

Pressure must also enter both the momentum and energy fluxes. Its spatial contribution must reproduce the isotropic stress $pg^{ij}$, while its components involving the fifth direction must account for pressure work. We therefore need a pressure tensor defined relative to the moving fluid.

To construct these additional contributions, we introduce the vector $\xi^A=\delta^A_4$ tangent to the auxiliary direction. Since $\xi^A$ has only one non-zero component, $\xi^4=1$, the symmetric combination $u^A\xi^B+\xi^Au^B$ contributes only when at least one index equals $4$. Multiplying it by $-\rho e$ introduces the internal-energy density and its transport with the fluid into the fifth row and column, without changing the mass and momentum components. We define the pressure tensor by
\begin{equation}
  \mathcal P^{AB}=g^{AB}-u^A\xi^B-\xi^Au^B.
  \label{eq:fluid_projector}
\end{equation}
Its components are
\begin{equation*}
\begin{aligned}
  \mathcal P^{0A}&=0,\qquad \mathcal P^{ij}=g^{ij},\\
  \mathcal P^{i4}&=g^{i4}-u^i,\qquad
  \mathcal P^{44}=g^{44}-2u^4.
\end{aligned}
\end{equation*}
Thus, $\mathcal P^{ij}=g^{ij}$ reproduces the isotropic spatial pressure stress, while the components involving index $4$ account for pressure work without changing the mass density or mass flux.

The symmetric \emph{mass--momentum tensor} for the inviscid fluid is
\begin{equation}
T^{AB}
=\rho u^Au^B-\rho e(u^A\xi^B+\xi^Au^B)
+p\,\mathcal P^{AB}.
\label{eq:T5D}    
\end{equation}
Equivalently,
\begin{equation}
  T^{AB}=\rho u^Au^B+pg^{AB}
  -(\rho e+p)(u^A\xi^B+\xi^Au^B).
  \label{eq:T5D_expanded}
\end{equation}
This is the ideal null-fluid constitutive structure of \citet{banerjee2016null}, written with the convention $u_A\xi^A=1$. The internal-energy term and the components of the pressure tensor involving index $4$ do not alter $T^{00}$, $T^{0i}$, or $T^{ij}$. They instead give
\begin{equation}
\begin{aligned}
  T^{04}=T^{40}&=\rho(u^4-e),\\
  T^{i4}=T^{4i}&=\rho(u^4-e)u^i+p(g^{i4}-u^i),\\
  T^{44}&=\rho(u^4)^2-2\rho e\,u^4+p(g^{44}-2u^4).
\end{aligned}
\label{eq:T_fibre}
\end{equation}
In a flat frame with $g_{0i}=0$ and $\Phi=0$, these entries reduce to
\begin{equation*}
  T^{04}=-\rho(e+K),\qquad
  T^{i4}=-[\rho(e+K)+p]u^i,\qquad
  K=\tfrac12g_{ij}u^iu^j.
\end{equation*}
The fifth column carries the negative total-energy density and flux in that frame, including pressure work, with the minus sign following from the positive cross term between $t$ and $x^4$ in the five-dimensional metric.

The geometric properties of this construction follow from the metric and velocity identities,
\begin{equation*}
  \xi_A=\delta_A^0,\qquad
  \xi_A\xi^A=0,\qquad
  u_A\xi^A=1,\qquad
  \nabla_A\xi^B=0.
\end{equation*}
The tensor $\mathcal P^{AB}$ satisfies $\mathcal P^{AB}u_B=\mathcal P^{AB}\xi_B=0$ and therefore restricts the pressure stress to directions transverse to both the fluid velocity and the null auxiliary direction.

For the inviscid adiabatic flow considered here, the five balance equations determine $\rho$, the three spatial velocities, and $e$, with pressure supplied by the equation of state. Viscosity and heat conduction would require additional constitutive terms, including their energy fluxes; these extensions are not developed here.

\subsection{The master balance equation}

The fluid equations are written as the single covariant statement
\begin{equation}
  \nabla_A T^{AB}= \frac{1}{\Gdet}
  \frac{\partial}{\partial x^A}\left(\Gdet\,T^{AB}\right)
  + \Gamma^B_{AC}T^{AC}
  = 0,
  \label{eq:master}
\end{equation}
using the covariant divergence formula \eqref{eq:divT}. The five values of the free index $B = 0,1,2,3,4$ correspond to five partial differential equations: $B=0$ gives the mass continuity equation, $B=1,2,3$ give the momentum balance equations, and $B=4$, together with continuity and momentum, gives the internal-energy equation.

\paragraph{Continuity equation}
For $B=0$, $T^{A0}=\rho u^A-\rho e\xi^A$. The extra term has only a component with index $4$, whose derivative vanishes. Together with $\Gamma^0_{AB}=0$, this gives
\begin{equation}
  \frac{\partial}{\partial x^\mu}\left(\Gdet\,\rho u^\mu\right)=0,
  \label{eq:continuity}
\end{equation}
where $\mu=0,1,2,3$. Setting $u^0=1$, this may be written in the flux form
\begin{equation}
  \frac{\partial}{\partial t}\left(\Gdet\,\rho\right)
  + \frac{\partial}{\partial x^i}\left(\Gdet\,\rho u^i\right)
  =0.
  \label{eq:continuity_flux}
\end{equation}
The factor $\Gdet$ accounts for the coordinate volume element. No derivative with respect to $x^4$ appears because all physical fields are independent of $x^4$.

\paragraph{Momentum equations}

For $B=i$, we substitute $T^{0i}=\rho u^i$ and $T^{ji}=\rho u^ju^i+pg^{ji}$ into \eqref{eq:master}. All derivatives with respect to $x^4$ and all Christoffel symbols with a lower index $4$ vanish. Combining the pressure-flux divergence with the pressure contribution from the connection gives
\begin{equation*}
  \frac{1}{\Gdet}\frac{\partial}{\partial x^j}\left(\Gdet\,pg^{ji}\right)
  +p\Gamma^i_{jk}g^{jk}
  =g^{ij}\frac{\partial p}{\partial x^j}.
\end{equation*}
The momentum equations therefore reduce to
\begin{equation}
\begin{aligned}
  &\frac{\partial}{\partial t}(\Gdet\,\rho u^i)
  +\frac{\partial}{\partial x^j}(\Gdet\,\rho u^ju^i)\\
  &\qquad=-\Gdet\left[
    g^{ij}\frac{\partial p}{\partial x^j}
    +\rho\left(\Gamma^i_{00}+2\Gamma^i_{0j}u^j
      +\Gamma^i_{jk}u^ju^k\right)
  \right].
  \label{eq:momentum_flux}
\end{aligned}
\end{equation}
The left-hand side describes the local change and transport of momentum. The right-hand side contains the pressure gradient and the geometric contributions, identified as gravity, centrifugal acceleration, Coriolis acceleration, and spatial-coordinate terms in the rotating spherical example in Section~\ref{sec:sphere}.

\paragraph{Energy and thermodynamics}

The fifth component of the master equation completes the mass and momentum balances with an equation involving energy. Substituting the components $T^{04}$ and $T^{j4}$ from \eqref{eq:T_fibre} gives
\begin{equation}
\begin{aligned}
  &\frac{\partial}{\partial t}\left[\Gdet\,\rho(u^4-e)\right]\\
  &\quad+\frac{\partial}{\partial x^j}\left\{\Gdet
    \left[\rho(u^4-e)u^j+p(g^{j4}-u^j)\right]\right\}
  =-\Gdet\,\Gamma^4_{AC}T^{AC}.
\end{aligned}
\label{eq:energy_fibre}
\end{equation}
For smooth solutions, substituting $u^4=-L$ and using continuity and the momentum equations cancels the mechanical-energy terms. The remaining terms give the internal-energy balance in flux form:
\begin{equation}
\begin{aligned}
  &\frac{\partial}{\partial t}(\Gdet\,\rho e)
  +\frac{\partial}{\partial x^i}(\Gdet\,\rho e\,u^i)\\
  &\qquad=-p\left[
    \frac{\partial\Gdet}{\partial t}
    +\frac{\partial}{\partial x^i}(\Gdet\,u^i)
  \right].
\end{aligned}
  \label{eq:internal_energy_flux}
\end{equation}
The left-hand side describes the local change and transport of internal energy. The right-hand side accounts for pressure work, which decreases internal energy during expansion and increases it during compression. The time derivative of the volume factor retains the effect of moving coordinates.

For smooth solutions, the master equation is equivalent to continuity, the three momentum equations, and the internal-energy equation. Its integral form, suitable for conservative discretization, is derived in Appendix~\ref{sec:gauss}.

\section{Different Formulations of the Equations}
\label{sec:forms}

The balance laws derived above describe densities and their fluxes. For smooth solutions, they can also be written in terms of changes along fluid trajectories or expressed using different prognostic variables. We first derive alternative forms of continuity, then consider the advective, covariant flux, Lamb, and Crocco forms of momentum. The final subsection discusses thermodynamic variables and total energy. 

\subsection{Continuity equation}

The advective form describes the change in density along a fluid trajectory. Expanding \eqref{eq:continuity} by the product rule,
\begin{equation*}
  \Gdet\,u^\mu\frac{\partial\rho}{\partial x^\mu}
  +
  \rho\,\frac{\partial}{\partial x^\mu}\left(\Gdet\,u^\mu\right)
  =0,
\end{equation*}
yields the advective form
\begin{equation}
  \frac{\mathrm D\rho}{\mathrm Dt}
  =
  -\rho\,
  \frac{1}{\Gdet}
  \frac{\partial}{\partial x^\mu}\left(\Gdet\,u^\mu\right),
  \label{eq:continuity_advective}
\end{equation}
or equivalently
\begin{equation}
  \frac{\mathrm D\ln\rho}{\mathrm Dt}
  =
  -\frac{1}{\Gdet}
  \frac{\partial}{\partial x^\mu}\left(\Gdet\,u^\mu\right).
  \label{eq:continuity_advective_compact}
\end{equation}
The intrinsic derivative acting on the scalars $\rho$ or $\ln\rho$ reduces to the ordinary derivative along the trajectory,
\begin{equation*}
  \frac{\mathrm D}{\mathrm Dt}
  =
  u^\mu\frac{\partial}{\partial x^\mu}
  =
  \frac{\partial}{\partial t}
  +u^i\frac{\partial}{\partial x^i}.
\end{equation*}

Continuity can also be expressed using a prognostic variable other than density. For a dry ideal gas, the equation of state relates density to the Exner function $\pi$ and potential temperature $\theta$ through
\begin{equation}
  \pi
  \equiv
  \left(\frac{p}{p_0}\right)^{R_d/c_{pd}}
  =
  \left(\frac{R_d\rho\theta}{p_0}\right)^{\gamma_d-1},
  \qquad
  T=\theta\pi ,
  \label{eq:exner}
\end{equation}
where $\theta$ is the potential temperature, $\gamma_d=c_{pd}/c_{vd}$ and
\begin{equation*}
  \frac{R_d}{c_{pd}}
  =
  \frac{\gamma_d-1}{\gamma_d}.
\end{equation*}
Equation~\eqref{eq:exner} may also be written
\begin{equation*}
  \rho
  =
  \frac{p_0}{R_d\theta}
  \pi^{1/(\gamma_d-1)}.
\end{equation*}
Taking its material derivative and using
\eqref{eq:continuity_advective_compact} gives
\begin{equation}
  \frac{1}{\gamma_d-1}
  \frac{1}{\pi}\frac{\mathrm D\pi}{\mathrm Dt}
  -
  \frac{1}{\theta}\frac{\mathrm D\theta}{\mathrm Dt}
  +
  \frac{1}{\Gdet}
  \frac{\partial}{\partial x^\mu}\left(\Gdet\,u^\mu\right)
  =0.
  \label{eq:continuity_exner_general}
\end{equation}
For a dry adiabatic flow, potential temperature is conserved along trajectories, $\mathrm D\theta/\mathrm Dt=0$, as derived in Section~\ref{subsec:thermodynamics}. The Exner equation then becomes
\begin{equation}
  \frac{\mathrm D\pi}{\mathrm Dt}
  =
  -(\gamma_d-1)\pi\,
  \frac{1}{\Gdet}
  \frac{\partial}{\partial x^\mu}\left(\Gdet\,u^\mu\right).
  \label{eq:exner_advective}
\end{equation}

\subsection{Momentum equation}

\subsubsection{Advective form}

The advective form expresses the acceleration of a fluid parcel. Expanding the momentum flux in \eqref{eq:momentum_flux} by the product rule gives
\begin{equation}
  u^i
  \left[
    \frac{1}{\Gdet}
    \frac{\partial}{\partial x^\mu}\left(\Gdet\,\rho u^\mu\right)
  \right]
  +
  \rho\,u^A\nabla_A u^i
  +
  g^{ij}\frac{\partial p}{\partial x^j}
  =0.
\end{equation}
The term in brackets is the continuity equation and therefore vanishes. The momentum equation becomes
\begin{equation}
  \frac{\mathrm D u^i}{\mathrm Dt} = \frac{\mathrm d u^i}{\mathrm dt}
  +
  \Gamma^i_{AB}u^Au^B
  =-\frac{g^{ij}}{\rho}\frac{\partial p}{\partial x^j},
  \label{eq:momentum_advective_contra}
\end{equation}
where
\begin{equation}
  \frac{\mathrm d u^i}{\mathrm dt}
  =
  \frac{\partial u^i}{\partial t}
  +
  u^j\frac{\partial u^i}{\partial x^j}.
\end{equation}
The ordinary derivative describes changes in the coordinate components of velocity. The connection term completes the intrinsic acceleration and includes gravity, rotation, and the effects of curvilinear coordinates.

An advective form of the momentum equation can also be written for the covariant velocity components $u_i$. Because the Levi--Civita connection is metric-compatible, lowering the index commutes with covariant differentiation. The intrinsic acceleration can then be written directly in covariant form:
\begin{equation}
  \frac{\mathrm D u_i}{\mathrm Dt}
  =
  -\frac{1}{\rho}\,\frac{\partial p}{\partial x^i},
  \label{eq:momentum_advective_cov}
\end{equation}
with
\begin{equation*}
  \frac{\mathrm D u_i}{\mathrm Dt}
  \equiv
  u^A\nabla_A u_i,
  \qquad
  u_i=g_{ij}u^j+g_{0i}.
\end{equation*}

\subsubsection{Covariant flux form}

The contravariant momentum balance \eqref{eq:momentum_flux} contains connection coefficients in its geometric source. Lowering the momentum index gives a flux form in which this source is expressed through derivatives of the metric. We define
\begin{equation}
  T^A{}_i
  \equiv
  g_{iC}T^{AC}
  =
  \rho u^A u_i+p\,\delta^A{}_i-(\rho e+p)\xi^A u_i.
  \label{eq:Tmixed_forms}
\end{equation}
The last term contributes only when $A=4$, so it does not change the time density or spatial momentum flux. Its ordinary divergence vanishes by independence from $x^4$, and its connection contribution vanishes because $\Gamma^C_{4i}=0$. The mixed-tensor divergence is
\begin{equation}
  \nabla_A T^A{}_i
  =
  \frac{1}{\Gdet}
  \frac{\partial}{\partial x^A}\left(\Gdet\,T^A{}_i\right)
  -
  \Gamma^C_{Ai}T^A{}_C.
\end{equation}
Using the symmetry of $T^{AB}$ and the definition of the Levi--Civita connection,
\begin{equation}
  \Gamma^C_{Ai}T^A{}_C
  =
  \frac{1}{2}T^{AC}\frac{\partial g_{AC}}{\partial x^i},
\end{equation}
so the covariant momentum flux equation becomes
\begin{equation}
  \frac{\partial}{\partial t}\left(\Gdet\,\rho u_i\right)
  +
  \frac{\partial}{\partial x^j}\left[
    \Gdet\left(
      \rho u^ju_i+p\,\delta^j{}_i
    \right)
  \right]
  =
  \frac{1}{2}\Gdet\,T^{AC}\frac{\partial g_{AC}}{\partial x^i},
  \label{eq:momentum_flux_cov_forms}
\end{equation}
where the $A=4$ derivative vanishes because all physical fields are independent of $x^4$.

In \eqref{eq:momentum_flux_cov_forms}, the geometric source term is expressed entirely in terms of spatial derivatives of the metric tensor. Expanding it gives
\begin{equation}
  \frac{1}{2}T^{AC}\frac{\partial g_{AC}}{\partial x^i}
  =
  \rho\left(
    -\frac{\partial\Phie}{\partial x^i}
    +u^j\frac{\partial g_{0j}}{\partial x^i}
    +\frac{1}{2}u^ju^k\frac{\partial g_{jk}}{\partial x^i}
  \right)
  +
  p\,\frac{\partial\ln\Gdet}{\partial x^i} ,
  \label{eq:metric_source_forms}
\end{equation}
so that the gravitational, rotational, and curvilinear-coordinate contributions all arise from the spatial gradient of $g_{AC}$.

\subsubsection{Lamb form}

The Lamb form \citep{lamb1945hydrodynamics, hirsch2007numerical}, often called the \emph{vector-invariant form} in atmospheric dynamics \citep{ringler2011momentum}, separates the momentum equation into a mechanical-energy gradient, a vorticity term, and the pressure gradient. To obtain it from \eqref{eq:momentum_advective_cov}, we rewrite the intrinsic acceleration using an antisymmetric derivative of $u_A$. The required identity follows by differentiating the null condition $u^Au_A=0$:
\begin{equation*}
  \nabla_i\left(u^A u_A\right)
  =
  u_A\nabla_i u^A
  +
  u^A\nabla_i u_A
  =0.
\end{equation*}
Metric compatibility implies $u_A\nabla_i u^A=u^A\nabla_i u_A$. The differentiated null condition therefore gives $2u^A\nabla_i u_A=0$. This vanishing contraction can be subtracted from the intrinsic acceleration without changing its value:
\begin{equation*}
  \frac{\mathrm D u_i}{\mathrm Dt}
  =
  u^A\nabla_A u_i
  =
  u^A\left(\nabla_A u_i-\nabla_i u_A\right).
\end{equation*}
This antisymmetric combination is useful because the Levi--Civita connection is symmetric in its lower indices, so its Christoffel terms cancel:
\begin{equation}
  \nabla_A u_i-\nabla_i u_A
  =
  \frac{\partial u_i}{\partial x^A}-\frac{\partial u_A}{\partial x^i},
  \label{eq:lamb_christoffel_cancel}
\end{equation}
and we can write the intrinsic derivative as
\begin{equation*}
  \frac{\mathrm D u_i}{\mathrm Dt}
  =
  u^A\left(\frac{\partial u_i}{\partial x^A}-\frac{\partial u_A}{\partial x^i}\right).
\end{equation*}
Using $u^0=1$, $u_4=1$, the independence from $x^4$, and
\begin{equation*}
  u_0
  =
  -\Phie
  -\frac{1}{2}g_{jk}u^ju^k,
\end{equation*}
the momentum equation becomes
\begin{equation}
  \frac{\partial u_i}{\partial t}
  +
  \frac{\partial}{\partial x^i}
  \left(
    \frac{1}{2}g_{jk}u^ju^k+\Phie
  \right)
  =
  u^j\left(\frac{\partial u_j}{\partial x^i}-\frac{\partial u_i}{\partial x^j}\right)
  -
  \frac{1}{\rho}\frac{\partial p}{\partial x^i}.
  \label{eq:lamb_momentum}
\end{equation}
No Christoffel symbols appear explicitly in the Lamb form \eqref{eq:lamb_momentum}; the geometry and rotating-coordinate effects enter through the covariant velocity $u_i$, the effective potential $\Phie$, and their ordinary derivatives. 

\subsubsection{Crocco form}

The Crocco form follows from the Lamb form by combining the pressure gradient with the mechanical-energy gradient. For a dry ideal gas, the pressure term can be separated into enthalpy and entropy contributions. In terms of the Exner function, differentiating \eqref{eq:exner} and using $p=\rho R_dT$ gives
\begin{equation}
  \frac{1}{\rho}\frac{\partial p}{\partial x^i}
  =
  c_{pd}\theta\,
  \frac{\partial\pi}{\partial x^i}.
  \label{eq:exner_pressure_gradient}
\end{equation}
Since $T=\theta\pi$, \eqref{eq:exner_pressure_gradient} may also be written
\begin{equation}
  \frac{1}{\rho}\frac{\partial p}{\partial x^i}
  =
  \frac{\partial}{\partial x^i}
  \left(c_{pd}\theta\pi\right)
  -
  c_{pd}\pi\,
  \frac{\partial\theta}{\partial x^i}.
  \label{eq:exnergrad}
\end{equation}

The specific enthalpy and entropy of a dry ideal gas may be written
\begin{equation*}
  h=c_{pd}T=c_{pd}\theta\pi,
  \qquad
  s=c_{pd}\ln\theta+\mathrm{const.},
\end{equation*}
so that \eqref{eq:exnergrad} is the Gibbs relation
\begin{equation*}
  \mathrm dh
  =
  T\,\mathrm ds
  +
  \frac{1}{\rho}\,\mathrm dp
\end{equation*}
written in terms of $\theta$ and $\pi$.

Substitution of \eqref{eq:exnergrad} into the Lamb equation
\eqref{eq:lamb_momentum} gives the Crocco form
\begin{equation}
  \frac{\partial u_i}{\partial t}
  +
  \frac{\partial}{\partial x^i}
  \left(
    \frac{1}{2}g_{jk}u^ju^k
    +\Phie
    + c_{pd}\theta\pi
  \right)
  =
  u^j\left(\frac{\partial u_j}{\partial x^i}-\frac{\partial u_i}{\partial x^j}\right)
  +
  c_{pd}\pi\,\frac{\partial\theta}{\partial x^i}.
  \label{eq:crocco}
\end{equation}
The Crocco form separates the momentum equation into gradients of mechanical energy and enthalpy, together with vorticity and entropy-gradient terms \citep{hirsch2007numerical}.

\subsubsection{Lamb and Crocco forms with coordinate-relative velocity}
\label{subsubsec:relative_forms}

Through $g_{0i}$, the covariant velocity $u_i=g_{ij}u^j+g_{0i}$ contains a contribution fixed by the metric. In a rotating frame, this contribution may be substantially larger than the atmospheric velocity (Section~\ref{sec:sphere}). A prognostic equation for $u_i$ therefore carries a large, known background component in addition to the flow-dependent part. For numerical calculations, it can be advantageous to separate these contributions. In rotating coordinates, the separation is also useful physically, because it distinguishes the relative and planetary contributions to the vorticity term.

To separate the background contribution, we define
\begin{equation}
  v_i
  \equiv
  g_{ij}u^j,
  \qquad
  u_i=v_i+g_{0i}.
  \label{eq:v_def}
\end{equation}
To interpret these variables, consider a reference motion $\bar u^A$, with $\bar u^0=1$, orthogonal to the spatial coordinate directions. Since $g_{i4}=0$, the condition $\bar u_i=0$ gives
\begin{equation*}
  g_{0i}+g_{ij}\bar u^j=0,
  \qquad \bar u^i=-g^{ij}g_{0j}.
\end{equation*}
In rotating spherical coordinates, $\bar u^1=-\Omega$, corresponding to rest in the non-rotating frame. Lowering the fluid velocity relative to this reference motion with the spatial metric gives
\begin{equation*}
  g_{ij}\bigl(u^j+g^{jk}g_{0k}\bigr)
  =
  u_i,
\end{equation*}
whereas $v_i$ is obtained in the same way from the velocity relative to the coordinates. The contravariant velocity follows from $u^i=g^{ij}v_j$, and the kinetic energy is $\frac12g_{jk}u^ju^k=\frac12g^{jk}v_jv_k$.

With \eqref{eq:v_def}, $\partial u_i/\partial t=\partial v_i/\partial t+\partial g_{0i}/\partial t$, and the antisymmetric combination in the Lamb form separates as
\begin{equation}
  \frac{\partial u_j}{\partial x^i}-\frac{\partial u_i}{\partial x^j}
  =
  \left(\frac{\partial v_j}{\partial x^i}-\frac{\partial v_i}{\partial x^j}\right)
  +
  \left(\frac{\partial g_{0j}}{\partial x^i}-\frac{\partial g_{0i}}{\partial x^j}\right).
  \label{eq:vorticity_split}
\end{equation}
Substituting into the Lamb form \eqref{eq:lamb_momentum} gives its coordinate-relative variant,
\begin{equation}
  \begin{aligned}
  \frac{\partial v_i}{\partial t}
  +
  \frac{\partial}{\partial x^i}
  \left(
    \frac{1}{2}g_{jk}u^ju^k+\Phie
  \right)
  ={}&
  u^j\left(\frac{\partial v_j}{\partial x^i}-\frac{\partial v_i}{\partial x^j}\right)
  +
  u^j\left(\frac{\partial g_{0j}}{\partial x^i}-\frac{\partial g_{0i}}{\partial x^j}\right)
  \\
  & -\frac{\partial g_{0i}}{\partial t}
  -
  \frac{1}{\rho}\frac{\partial p}{\partial x^i},
  \end{aligned}
  \label{eq:lamb_relative}
\end{equation}
and substituting into the Crocco form \eqref{eq:crocco} gives its coordinate-relative variant,
\begin{equation}
  \begin{aligned}
  \frac{\partial v_i}{\partial t}
  +
  \frac{\partial}{\partial x^i}
  \left(
    \frac{1}{2}g_{jk}u^ju^k
    +\Phie
    + c_{pd}\theta\pi
  \right)
  ={}&
  u^j\left(\frac{\partial v_j}{\partial x^i}-\frac{\partial v_i}{\partial x^j}\right)
  +
  u^j\left(\frac{\partial g_{0j}}{\partial x^i}-\frac{\partial g_{0i}}{\partial x^j}\right)
  \\
  &-
  \frac{\partial g_{0i}}{\partial t}
  +
  c_{pd}\pi\,\frac{\partial\theta}{\partial x^i}.
  \end{aligned}
  \label{eq:crocco_relative}
\end{equation}
These equations only regroup the terms of \eqref{eq:lamb_momentum} and \eqref{eq:crocco}. Like the original forms, they contain no Christoffel symbols. The frame motion enters through the antisymmetric derivatives $\partial g_{0j}/\partial x^i-\partial g_{0i}/\partial x^j$, which are known from the metric, and through $\partial g_{0i}/\partial t$, which vanishes for a time-independent metric. The identity $u^iu^j\left(\partial g_{0j}/\partial x^i-\partial g_{0i}/\partial x^j\right)=0$ implies that the corresponding term does no work, as is also the case for the term $u^j\left(\partial v_j/\partial x^i-\partial v_i/\partial x^j\right)$. In a rotating frame, this term corresponds to the Coriolis acceleration.

As a generalization, any prescribed $\alpha_i(x^\mu)$, independent of $x^4$, may be subtracted from $u_i$; the derivation is unchanged, with $g_{0i}$ replaced by $\alpha_i$ in \eqref{eq:vorticity_split}--\eqref{eq:crocco_relative} and $v_i$ by $u_i-\alpha_i$. This can be useful, for example, when the coordinate surfaces move and $g_{0i}$ also contains contributions from that motion.

\subsection{Thermodynamic equation}
\label{subsec:thermodynamics}

The internal-energy balance \eqref{eq:internal_energy_flux} can be expressed using the thermodynamic variables commonly chosen in atmospheric models. We consider a dry ideal gas with constant heat capacities, $e=c_{vd}T$ and $p=\rho R_dT$. Total energy requires a separate discussion because its conservation is related to time independence of the metric.

\subsubsection{Thermodynamic prognostic variables}

Combining \eqref{eq:internal_energy_flux} with continuity gives conservation of potential temperature along fluid trajectories,
\begin{equation}
  \frac{\mathrm D\theta}{\mathrm Dt}=0.
  \label{eq:theta_advective}
\end{equation}
Using the continuity equation, the same relation may be written in flux form as
\begin{equation}
  \frac{\partial}{\partial t}\left(\Gdet\,\rho\theta\right)
  +
  \frac{\partial}{\partial x^i}\left(\Gdet\,\rho u^i\theta\right)
  =0.
  \label{eq:theta_flux}
\end{equation}

Equivalent advective equations may be obtained for other thermodynamic variables. From $T=\theta\pi$ and
\eqref{eq:exner_advective}, an adiabatic flow satisfies
\begin{equation}
  \frac{\mathrm DT}{\mathrm Dt}
  +
  (\gamma_d-1)T\,\nabla_Au^A
  =0,
  \label{eq:temperature_advective}
\end{equation}
while the pressure satisfies
\begin{equation}
  \frac{\mathrm Dp}{\mathrm Dt}
  +
  \gamma_d p\,\nabla_Au^A
  =0.
  \label{eq:pressure_advective}
\end{equation}

\subsubsection{Total-energy balance}

Total energy combines internal energy with the mechanical energy associated with the chosen coordinates. We define the specific total energy by
\begin{equation*}
  E=e+E_{\mathrm{mech}}
   =e+\frac12g_{ij}u^iu^j+\Phie.
\end{equation*}
Its balance follows from the master equation by lowering the free index and selecting the time component, $\nabla_A T^A{}_0=0$. Substituting \eqref{eq:T5D_expanded} gives
\begin{equation*}
  T^0{}_0=-\rho E,\qquad
  T^i{}_0=-(\rho E+p)u^i.
\end{equation*}
Thus, $-T^0{}_0$ is the total-energy density and $-T^i{}_0$ is its spatial flux. The flux includes both energy transported by the fluid and work done by pressure forces. Potential energy enters through the metric used to lower the index.

As in the covariant momentum flux form \eqref{eq:momentum_flux_cov_forms}, the connection contribution can be expressed through derivatives of the metric. For the time component, the mixed-tensor divergence gives
\begin{equation}
  \frac{\partial}{\partial x^A}(\Gdet\,T^A{}_0)
  =\frac12\Gdet\,T^{AB}\frac{\partial g_{AB}}{\partial t}.
  \label{eq:energy_metric_source}
\end{equation}
Since all fields are independent of $x^4$, only the time and spatial derivatives remain on the left-hand side. Substituting the density and flux above therefore yields
\begin{equation}
  \frac{\partial}{\partial t}(\Gdet\,\rho E)
  +\frac{\partial}{\partial x^i}\left[\Gdet\,(\rho E+p)u^i\right]
  =-\frac12\Gdet\,T^{AB}\frac{\partial g_{AB}}{\partial t}.
  \label{eq:total_energy_time_dependent}
\end{equation}
The right-hand side accounts for the time dependence of the metric, including that introduced by moving coordinates. It describes changes in the energy associated with the chosen coordinates; a time-dependent coordinate transformation does not itself imply physical energy creation or loss.

For a stationary metric, $\partial g_{AB}/\partial t=0$, the source vanishes and the balance reduces to total-energy conservation:
\begin{equation}
  \frac{\partial}{\partial t}(\Gdet\,\rho E)
  +\frac{\partial}{\partial x^i}\left[\Gdet\,(\rho E+p)u^i\right]=0.
  \label{eq:total_energy_flux}
\end{equation}
This includes the uniformly rotating spherical geometry considered below.

\section{Rotating Spherical Coordinates}
\label{sec:sphere}

We next apply the formulation to an atmosphere described by longitude, latitude, and radius in a frame rotating with the planet. We construct the metric and its connection and recover the well-known advective equations following the rotating-coordinate example of \citet{Charron2014}. Gravity is included in the metric rather than added as an external force. The full radial dependence and all rotational terms are retained in this section.

\subsection{Distances and the rotating metric}

Let $(x^1,x^2,x^3)=(\lambda,\varphi,r)$, where $\lambda$ is longitude in the rotating frame, $\varphi$ is geographic latitude, and $r$ is distance from the planet's centre. An infinitesimal eastward displacement has length $r\cos\varphi\,\mathrm d\lambda$, whereas a northward displacement has length $r\,\mathrm d\varphi$. The spatial distance and volume elements are
\begin{equation*}
  \mathrm d\ell^2=r^2\cos^2\varphi\,\mathrm d\lambda^2
                 +r^2\,\mathrm d\varphi^2+\mathrm dr^2,
  \qquad
  \mathrm dV=r^2\cos\varphi\,\mathrm d\lambda\,\mathrm d\varphi\,\mathrm dr.
\end{equation*}
These scale factors give $g_{ij}=\mathrm{diag}(r^2\cos^2\varphi,r^2,1)$ and $\Gdet=r^2\cos\varphi$.

For a frame rotating at constant angular velocity $\Omega$, the inertial longitude is $\lambda+\Omega t$. We replace its differential by $\mathrm d\lambda+\Omega\,\mathrm dt$ in the inertial line element. With the central gravitational potential $\Phi=-GM/r$, where $G$ is the Newtonian gravitational constant and $M$ is the planetary mass, this gives
\begin{equation}
  \mathrm ds^2
  =r^2\cos^2\varphi\,(\mathrm d\lambda+\Omega\,\mathrm dt)^2
   +r^2\,\mathrm d\varphi^2+\mathrm dr^2
   -2\Phi\,\mathrm dt^2+2\,\mathrm dt\,\mathrm dx^4.
  \label{eq:rotating_line_element}
\end{equation}
Expanding the square shows two effects of rotation. The cross term gives $g_{01}$, while the term proportional to $\mathrm dt^2$ adds the centrifugal potential:
\begin{equation}
  g_{01}=\Omega r^2\cos^2\varphi,\qquad g_{02}=g_{03}=0,
  \qquad
  \Phie=-\frac{GM}{r}-\frac12\Omega^2r^2\cos^2\varphi,
  \label{eq:A_Phi_sphere}
\end{equation}
with $g_{00}=-2\Phie$. The effective potential $\Phie$ includes both gravitational and centrifugal contributions, so its gradient accounts for both in the momentum equations. The remaining components involving index $4$ are unchanged: $g_{04}=1$ and $g_{i4}=g_{44}=0$.

\subsection{The spherical connection}

Differentiating the metric using \eqref{eq:christoffel} gives the connection coefficients below. These coefficients determine how gravity, rotation, and spherical curvature enter the momentum equations. In the coordinate acceleration $\Gamma^i_{AB}u^Au^B$, the $00$ entry gives the velocity-independent acceleration, the $0j$ entries supply the terms linear in velocity, and the spatial entries supply the curvature terms. Each off-diagonal entry contributes twice because both index orders occur in the sum.

Rows and columns are ordered $(t,\lambda,\varphi,r,x^4)$. Entries involving a lower index $4$ vanish by \eqref{eq:Gamma_lower5}. The spatial coefficients agree with \citet[Eqs.~61--63]{Charron2014}, except that $\Gamma^3_{00}$ here also contains $GM/r^2$, because gravity is included in the metric:
\begin{equation}
  \Gamma^1_{AB} =
  \begin{pmatrix}
    0 & 0 & -\Omega\tan\varphi & \Omega r^{-1} & 0 \\
    0 & 0 & -\tan\varphi & r^{-1} & 0 \\
    -\Omega\tan\varphi & -\tan\varphi & 0 & 0 & 0 \\
    \Omega r^{-1} & r^{-1} & 0 & 0 & 0 \\
    0 & 0 & 0 & 0 & 0
  \end{pmatrix},
  \label{eq:Gamma1}
\end{equation}
\begin{equation}
  \Gamma^2_{AB} =
  \begin{pmatrix}
    \Omega^2\cos\varphi\sin\varphi & \Omega\cos\varphi\sin\varphi & 0 & 0 & 0 \\
    \Omega\cos\varphi\sin\varphi & \cos\varphi\sin\varphi & 0 & 0 & 0 \\
    0 & 0 & 0 & r^{-1} & 0 \\
    0 & 0 & r^{-1} & 0 & 0 \\
    0 & 0 & 0 & 0 & 0
  \end{pmatrix},
  \label{eq:Gamma2}
\end{equation}
\begin{equation}
  \Gamma^3_{AB} =
  \begin{pmatrix}
    \dfrac{GM}{r^2}-r\Omega^2\cos^2\!\varphi & -r\Omega\cos^2\!\varphi & 0 & 0 & 0 \\
    -r\Omega\cos^2\!\varphi & -r\cos^2\!\varphi & 0 & 0 & 0 \\
    0 & 0 & -r & 0 & 0 \\
    0 & 0 & 0  & 0 & 0 \\
    0 & 0 & 0 & 0 & 0
  \end{pmatrix}.
  \label{eq:Gamma3}
\end{equation}

For the auxiliary direction, substituting the inverse metric from \eqref{eq:g_inv_block} into \eqref{eq:christoffel} gives only
\begin{equation}
  \Gamma^4_{03}=\Gamma^4_{30}=-\frac{GM}{r^2}.
  \label{eq:Gamma5_0r}
\end{equation}
All other upper-$4$ coefficients vanish. These coefficients contain only the gravitational contribution; rotation remains included through the other connection components and the effective potential.

\subsection{Recovering the advective equations}
\label{subsec:coriolis}

Combining the metric and connection in the spatial momentum balance gives
\begin{equation*}
  \frac{\mathrm du^i}{\mathrm dt}
  =-\frac{1}{\rho}g^{ij}\frac{\partial p}{\partial x^j}
   -\Gamma^i_{00}-2\Gamma^i_{0j}u^j-\Gamma^i_{jk}u^ju^k.
\end{equation*}
In these stationary rotating spherical coordinates, $-\Gamma^i_{00}$ represents gravity and centrifugal acceleration, $-2\Gamma^i_{0j}u^j$ supplies Coriolis acceleration, and $-\Gamma^i_{jk}u^ju^k$ contains the terms associated with the spatial coordinate basis. To recover the familiar equations for eastward, northward, and upward speeds, we must also differentiate the spherical scale factors relating these speeds to the coordinate velocities.

The components $u^1=\mathrm d\lambda/\mathrm dt$ and $u^2=\mathrm d\varphi/\mathrm dt$ are angular rates, not speeds. To interpret the equations in meteorological terms, define the eastward, northward, and upward velocities relative to the rotating frame by
\begin{equation*}
  U=r\cos\varphi\,u^1,\qquad V=r\,u^2,\qquad W=u^3.
\end{equation*}
The ordinary derivative along the trajectory, applied below to these individual components, is
\begin{equation*}
  \frac{\mathrm d}{\mathrm dt}
  =\frac{\partial}{\partial t}+\frac{U}{r\cos\varphi}\frac{\partial}{\partial\lambda}
             +\frac{V}{r}\frac{\partial}{\partial\varphi}+W\frac{\partial}{\partial r}.
\end{equation*}
For example, substituting the entries of \eqref{eq:Gamma1} into the longitude equation gives
\begin{equation*}
  \frac{\mathrm du^1}{\mathrm dt}
  +\frac{2}{r}(u^1+\Omega)u^3
  -2\tan\varphi\,(u^1+\Omega)u^2
  =-\frac{1}{\rho r^2\cos^2\varphi}\frac{\partial p}{\partial\lambda}.
\end{equation*}
The eastward speed is $U=r\cos\varphi\,u^1$. Differentiating this product, with $\mathrm dr/\mathrm dt=W$ and $\mathrm d\varphi/\mathrm dt=V/r$, yields
\begin{equation*}
  \frac{\mathrm dU}{\mathrm dt}
  =r\cos\varphi\,\frac{\mathrm du^1}{\mathrm dt}
   +\frac{UW}{r}-\frac{UV\tan\varphi}{r}.
\end{equation*}
Multiplying the longitude equation by $r\cos\varphi$ and using this identity gives the eastward equation below. Applying the same procedure to $V=ru^2$ and $W=u^3$ gives the northward and radial equations. We retain the derivatives of $\Phie$ to represent gravitational and centrifugal acceleration:

\begin{align}
  \frac{\mathrm dU}{\mathrm dt}
  -\frac{UV\tan\varphi}{r}+\frac{UW}{r}
  -2\Omega\sin\varphi\,V+2\Omega\cos\varphi\,W
  &=-\frac{1}{\rho r\cos\varphi}\frac{\partial p}{\partial\lambda},
  \label{eq:sphere_east}\\
  \frac{\mathrm dV}{\mathrm dt}
  +\frac{U^2\tan\varphi}{r}+\frac{VW}{r}
  +2\Omega\sin\varphi\,U
  &=-\frac{1}{\rho r}\frac{\partial p}{\partial\varphi}
    -\frac{1}{r}\frac{\partial\Phie}{\partial\varphi},
  \label{eq:sphere_north}\\
  \frac{\mathrm dW}{\mathrm dt}
  -\frac{U^2+V^2}{r}-2\Omega\cos\varphi\,U
  &=-\frac{1}{\rho}\frac{\partial p}{\partial r}-\frac{\partial\Phie}{\partial r}.
  \label{eq:sphere_up}
\end{align}
The terms quadratic in velocity and proportional to $1/r$ arise because the local eastward, northward, and upward directions vary with position. The terms proportional to $\Omega$ are the Coriolis terms. Those involving $2\Omega\sin\varphi$ couple the two horizontal velocities; those involving $2\Omega\cos\varphi$ couple eastward and vertical motion and are usually called the non-traditional terms.

For the effective potential in \eqref{eq:A_Phi_sphere}, these accelerations are
\begin{equation*}
  -\frac{1}{r}\frac{\partial\Phie}{\partial\varphi}
    =-\Omega^2r\cos\varphi\sin\varphi,\qquad
  -\frac{\partial\Phie}{\partial r}=-\frac{GM}{r^2}+\Omega^2r\cos^2\varphi.
\end{equation*}
The centrifugal acceleration points away from the rotation axis. Its projection onto the spherical basis has an outward radial component $\Omega^2r\cos^2\varphi$ and an equatorward meridional component $-\Omega^2r\cos\varphi\sin\varphi$. Both arise from the centrifugal contribution already included in $\Phie$; no separate centrifugal force is required. Thus effective gravity is not purely radial for the spherical reference surfaces and potential used here. Together with the Coriolis and spatial-coordinate terms displayed above, these expressions recover the familiar momentum equations in a rotating spherical frame.

The same volume factor gives the continuity equation,
\begin{equation*}
  \frac{\partial\rho}{\partial t}
  +\frac{1}{r\cos\varphi}\frac{\partial}{\partial\lambda}(\rho U)
  +\frac{1}{r\cos\varphi}\frac{\partial}{\partial\varphi}(\rho V\cos\varphi)
  +\frac{1}{r^2}\frac{\partial}{\partial r}(r^2\rho W)=0.
\end{equation*}

In terms of the physical velocity components, the energy is
\begin{equation*}
  E=e+\frac12(U^2+V^2+W^2)+\Phie.
\end{equation*}
It includes internal energy, kinetic energy relative to the rotating frame, and the full effective potential. Since this metric is stationary, the total energy above satisfies \eqref{eq:total_energy_flux}. Its spatial flux is $(\rho E+p)u^i$, including pressure work. The Coriolis terms do no work: multiplying their contributions in \eqref{eq:sphere_east}--\eqref{eq:sphere_up} by $U$, $V$, and $W$ makes them cancel in pairs.

\section{A Generalized Vertical Coordinate}
\label{sec:vertical}

Generalized vertical coordinates illustrate how tensor calculus recovers equations used in atmospheric models. Height-based terrain-following mappings \citep{gal1975use,schar2002new,klemp2011terrain} and hydrostatic-pressure coordinates \citep{laprise1992euler} are established constructions. Here we transform the metric for a specified mapping and recover the corresponding fluid equations.

\subsection{Coordinate mapping and metric}

Replace radius by a vertical label $\eta$, with $r=r(t,\lambda,\varphi,\eta)$ and $r_\eta\ne0$, leaving the other coordinates unchanged. Subscripts $t,\lambda,\varphi,\eta$ below denote partial derivatives at fixed values of the other new coordinates. Thus, $\mathrm dr=r_t\,\mathrm dt+r_\lambda\,\mathrm d\lambda+r_\varphi\,\mathrm d\varphi+r_\eta\,\mathrm d\eta$. Applying the metric transformation rule gives
\begin{equation}
  \hat g_{AB}=\frac{\partial x^C}{\partial\hat x^A}
              \frac{\partial x^D}{\partial\hat x^B}g_{CD},\qquad
  (\hat g_{ij})=\begin{pmatrix}
    r^2\cos^2\varphi+r_\lambda^2&r_\lambda r_\varphi&r_\lambda r_\eta\\
    r_\lambda r_\varphi&r^2+r_\varphi^2&r_\varphi r_\eta\\
    r_\lambda r_\eta&r_\varphi r_\eta&r_\eta^2
  \end{pmatrix}.
  \label{eq:gamma_hat}
\end{equation}
The time components and volume factor are
\begin{equation}
  \hat g_{00}=-2\Phie+r_t^2,\qquad
  \hat g_{0i}=\Omega r^2\cos^2\varphi\,\delta_i^1+r_t r_i,\qquad
  \mathcal J=|r_\eta|r^2\cos\varphi,
  \label{eq:eta_metric_time}
\end{equation}
with the auxiliary components unchanged. Here $r_i$ denotes $(r_\lambda,r_\varphi,r_\eta)$, and $\Phie$ is the original rotating-frame effective potential evaluated at the mapped radius. In particular, $-\hat g_{00}/2=\Phie-r_t^2/2$. Surface slopes introduce spatial cross terms, while coordinate motion also modifies the time components.

\subsection{Equations in the transformed coordinates}

Retain the physical velocities $U,V$ and $w=W$ of the rotating frame. The chain rule determines the velocity across coordinate surfaces and the material derivative:
\begin{equation}
\begin{aligned}
  \dot\eta&=\frac{1}{r_\eta}
    \left(w-r_t-\frac{U r_\lambda}{r\cos\varphi}-\frac{V r_\varphi}{r}\right),\\
  \frac{\mathrm D}{\mathrm Dt}&=\partial_t
    +\frac{U}{r\cos\varphi}\partial_\lambda
    +\frac{V}{r}\partial_\varphi+\dot\eta\,\partial_\eta.
\end{aligned}
  \label{eq:eta_velocity}
\end{equation}
Substituting $\mathcal J$ and the coordinate velocities into continuity gives
\begin{equation}
  \partial_t(\mathcal J\rho)
  +\partial_\lambda\left(\frac{\mathcal J\rho U}{r\cos\varphi}\right)
  +\partial_\varphi\left(\frac{\mathcal J\rho V}{r}\right)
  +\partial_\eta(\mathcal J\rho\dot\eta)=0.
  \label{eq:eta_continuity}
\end{equation}
The inverse spatial metric supplies the terrain-slope corrections to the pressure gradient. For example, $\hat g^{11}=(r\cos\varphi)^{-2}$ and $\hat g^{13}=-r_\lambda/[r_\eta(r\cos\varphi)^2]$ give
\begin{equation*}
  \hat g^{1j}p_j=\frac{1}{r^2\cos^2\varphi}
       \left(p_\lambda-\frac{r_\lambda}{r_\eta}p_\eta\right),\qquad
  \left.\partial_\lambda p\right|_r
       =p_\lambda-\frac{r_\lambda}{r_\eta}p_\eta.
\end{equation*}
Using the transformed connection and differentiating the scale factors relating coordinate velocities to $U,V,w$ yields
\begin{align}
  \frac{\mathrm DU}{\mathrm Dt}-\frac{UV\tan\varphi}{r}+\frac{Uw}{r}
  -2\Omega\sin\varphi\,V+2\Omega\cos\varphi\,w
  &=-\frac{p_\lambda-r_\lambda p_\eta/r_\eta}{\rho r\cos\varphi},
  \label{eq:eta_east}\\
  \frac{\mathrm DV}{\mathrm Dt}+\frac{U^2\tan\varphi}{r}+\frac{Vw}{r}
  +2\Omega\sin\varphi\,U
  &=-\frac{p_\varphi-r_\varphi p_\eta/r_\eta}{\rho r}
    -\frac{1}{r}\left.\partial_\varphi\Phie\right|_r,
  \label{eq:eta_north}\\
  \frac{\mathrm Dw}{\mathrm Dt}-\frac{U^2+V^2}{r}-2\Omega\cos\varphi\,U
  &=-\frac{p_\eta}{\rho r_\eta}-\partial_r\Phie.
  \label{eq:mom_r_eta}
\end{align}
The potential derivatives refer to the original rotating coordinates, with $\partial_r\Phie$ taken at fixed latitude. Coordinate motion enters through $\dot\eta$ and $\mathcal J$; $w$ remains the physical radial velocity. Potential-temperature advection \eqref{eq:theta_advective} and the equation of state complete the dry adiabatic system.

For a fixed terrain-following mapping $r=r_s+\eta(r_T-r_s)$, these equations give the spherical counterpart of the height-coordinate formulation of \citet{gal1975use}. More general surface shapes and motions enter through the same metric transformation; a pressure-based coordinate additionally requires its coordinate-defining relation.

\section{Thin-Shell and Spherical-Geopotential Approximations}
\label{sec:thinshell}

The generalized vertical mapping changes the coordinate representation while preserving the underlying geometry and dynamics. A geometric approximation, by contrast, specifies a new metric from which the connection must be recomputed. Deriving the connection directly from this approximate metric preserves the strict relationship between the geometry and the balance equations \citep{Charron2014}.

The approximation used below combines a thin-shell approximation to the horizontal geometry with a spherical effective potential and constant effective gravity \cite{Phillips1966}. The corresponding truncation of the time--space metric coefficients also yields the traditional form of the Coriolis terms. Specifically, we replace the radial coordinate $r$ by a constant reference radius $a$ in the horizontal spatial metric and the rotational time--space coefficients and approximate the effective potential by
\begin{equation*}
  \Phie^{(a)}=g_r z,\qquad z=r-a.
\end{equation*}
Here, $g_r$ is constant effective gravity (the subscript $r$ is a descriptive label, not a tensor index, and is not subject to Einstein summation). The coordinate $z$ is elevation above the reference sphere, on which the approximate potential is set to zero. This additional assumption replaces the full gravitational and centrifugal potential by a height-only effective potential; it is not merely a linearization of the gravitational potential. Using coordinates $(t,\lambda,\varphi,z,x^4)$, the approximate metric is
\begin{equation}
\begin{aligned}
  g^{(a)}_{11}&=a^2\cos^2\varphi,\qquad
  g^{(a)}_{22}=a^2,\qquad g^{(a)}_{33}=1,\\
  g^{(a)}_{01}&=\Omega a^2\cos^2\varphi,\qquad
  \Phie^{(a)}=g_r z,\\
  g^{(a)}_{00}&=-2g_r z,\qquad
  g^{(a)}_{04}=1,\qquad g^{(a)}_{44}=0.
\end{aligned}
\label{eq:shallow_metric}
\end{equation}
For this metric, the horizontal connection matrices are, with rows and columns ordered $(t,\lambda,\varphi,z,x^4)$,
\begin{equation}
  \Gamma^1_{AB} = -\tan\varphi
  \begin{pmatrix}
    0 & 0 & \Omega & 0 & 0\\
    0 & 0 & 1 & 0 & 0\\
    \Omega & 1 & 0 & 0 & 0\\
    0 & 0 & 0 & 0 & 0\\
    0 & 0 & 0 & 0 & 0
  \end{pmatrix},
  \qquad
  \Gamma^2_{AB} = \cos\varphi\sin\varphi
  \begin{pmatrix}
    0 & \Omega & 0 & 0 & 0\\
    \Omega & 1 & 0 & 0 & 0\\
    0 & 0 & 0 & 0 & 0\\
    0 & 0 & 0 & 0 & 0\\
    0 & 0 & 0 & 0 & 0
  \end{pmatrix}.
  \label{eq:thinshell_Gammas}
\end{equation}
In particular, $\Gamma^2_{00}=0$ because the approximate effective potential has no latitude dependence. The only non-zero upper-$3$ coefficient is $\Gamma^3_{00}=g_r$, which corresponds to the downward effective gravitational acceleration.
The non-zero upper-$4$ coefficients are
\begin{equation*}
  \Gamma^4_{02}=\Gamma^4_{20}=\Omega^2a^2\cos\varphi\sin\varphi,
  \qquad
  \Gamma^4_{03}=\Gamma^4_{30}=-g_r.
\end{equation*}
These coefficients enter the fifth balance equation; all other upper-$4$ coefficients vanish.

Here, the metric coefficients are independent of $t$ and $\lambda$, so the energy source in \eqref{eq:energy_metric_source} and the longitudinal momentum source in \eqref{eq:momentum_flux_cov_forms} both vanish. The approximate system therefore admits local conservation laws for total energy and axial angular momentum.

We can now recover the familiar atmospheric equations from this approximate geometry. We define the velocity components relative to the rotating frame and the Coriolis parameter by
\begin{equation*}
  U=a\cos\varphi\,u^1,\qquad V=a\,u^2,\qquad W=u^3,
  \qquad f=2\Omega\sin\varphi.
\end{equation*}
The material derivative is now
\begin{equation*}
  \frac{\mathrm d}{\mathrm dt}
  =\frac{\partial}{\partial t}
   +\frac{U}{a\cos\varphi}\frac{\partial}{\partial\lambda}
   +\frac{V}{a}\frac{\partial}{\partial\varphi}
   +W\frac{\partial}{\partial z}.
\end{equation*}
Substituting the approximate connection into \eqref{eq:momentum_advective_contra} and differentiating the horizontal scale factors gives
\begin{align}
  \frac{\mathrm dU}{\mathrm dt}
  -\frac{UV\tan\varphi}{a}-fV
  &=-\frac{1}{\rho a\cos\varphi}\frac{\partial p}{\partial\lambda},
  \label{eq:shallow_east}\\
  \frac{\mathrm dV}{\mathrm dt}
  +\frac{U^2\tan\varphi}{a}+fU
  &=-\frac{1}{\rho a}\frac{\partial p}{\partial\varphi},
  \label{eq:shallow_north}\\
  \frac{\mathrm dW}{\mathrm dt}
  &=-\frac{1}{\rho}\frac{\partial p}{\partial z}-g_r.
  \label{eq:shallow_up}
\end{align}
The volume factor $\sqrt{|\det g^{(a)}|}=a^2\cos\varphi$ gives continuity in the form
\begin{equation}
  \frac{\partial\rho}{\partial t}
  +\frac{1}{a\cos\varphi}\frac{\partial}{\partial\lambda}(\rho U)
  +\frac{1}{a\cos\varphi}\frac{\partial}{\partial\varphi}(\rho V\cos\varphi)
  +\frac{\partial}{\partial z}(\rho W)=0.
  \label{eq:shallow_continuity}
\end{equation}
For a dry adiabatic ideal gas, the system is completed by potential-temperature advection \eqref{eq:theta_advective}, with the material derivative given above, and the equation of state. In this way, the combined metric truncation and effective-potential approximation recover the standard nonhydrostatic shallow-atmosphere system with traditional Coriolis terms and constant effective gravity.

\section{Conclusion}
\label{sec:conclusion}

We presented an explicit five-dimensional formulation of compressible atmospheric dynamics in rotating spherical and generalized vertical coordinates, using the Eisenhart--Duval construction and the null-fluid mass--momentum tensor. The gravitational potential and the motion of the reference frame are included in the metric. We derived the geometric source terms and recovered the continuity, momentum, and internal-energy equations, together with their alternative forms. The total-energy balance accounts for time dependence of the metric and reduces to a conservation law for a stationary metric.

For generalized vertical coordinates, the transformed metric accounts for the slopes, stretching, and motion of the coordinate surfaces. The distinction between physical radial velocity and motion across those surfaces follows from the coordinate mapping. We also recovered the nonhydrostatic shallow-atmosphere equations with traditional Coriolis terms by combining a thin-shell metric with a spherical effective potential and constant effective gravity. These results establish an explicit connection between the five-dimensional geometric formulation and the equations standardly used in atmospheric models. The integral balance laws in Appendix~\ref{sec:gauss} provide a starting point for conservative discretization.

The equations considered in this paper describe a dry, inviscid gas. Future work will extend the formulation to viscosity, heat conduction, moisture, and the forcing terms supplied by physical parameterizations. We also plan to develop numerical methods that take advantage of this extended geometric formulation to preserve the balance laws.

\section*{Acknowledgements}

The authors thank Martin Charron for constructive comments and suggestions that helped improve the manuscript. S.T. gratefully acknowledges partial support for this research from the Pacific Institute for the Mathematical Sciences through its Collaborative Research Group on Structure-Preserving Discretizations and their Applications, the Natural Sciences and Engineering Research Council of Canada through its Alliance Advantage program, and Environment and Climate Change Canada through its Grants \& Contributions program. The funding from all three sources was awarded to R.J.S. A.N. acknowledges the support of the Natural Sciences and Engineering Research Council of Canada (NSERC), funding reference numbers RGPIN-2023-04121, RGPIN-2024-03851, and RGPIN-2020-05569. A.N. is also supported by Environment and Climate Change Canada, project number GCXE26S020.  S.S.S. is supported by NSERC (RGPIN-2020-05569) and Environment and Climate Change Canada (GCXE26S020).

\section*{Code Availability}

A Jupyter notebook that reproduces several of the results of this paper symbolically,
using SageManifolds, together with an illustration using Global Forecast System data, is publicly available at
\url{https://github.com/aradn97/geometric-geophysical-fluid-dynamics}.

\section*{Declaration of generative AI and AI-assisted technologies in the manuscript preparation process}
During the preparation of this work, the authors used Claude and Codex for copyediting and language refinement of the manuscript, and to assist in writing and
checking the code of the accompanying notebook. After using these tools, the authors reviewed and edited the content as needed and take full responsibility for the content of the published article and the accompanying code.

\bibliographystyle{apacite}
\bibliography{references}

@techreport{Avis1976,
  author      = {Avis, L. M.},
  title       = {A space--time tensor formulation for continuum mechanics in
                 general curvilinear, moving, and deforming coordinate systems},
  institution = {NASA Langley Research Center},
  year        = {1976},
  number      = {NASA-TR-R-462},
  address     = {Hampton, VA}
}

@inproceedings{cartan1923varietes,
  title={Sur les vari{\'e}t{\'e}s {\`a} connexion affine et la th{\'e}orie de la relativit{\'e} g{\'e}n{\'e}ralis{\'e}e (premi{\`e}re partie)},
  author={Cartan, {\'E}lie},
  booktitle={Annales scientifiques de l'{\'E}cole normale sup{\'e}rieure},
  volume={40},
  pages={325--412},
  year={1923}
}

@inproceedings{cartan1924varietes,
  title={Sur les vari{\'e}t{\'e}s {\`a} connexion affine, et la th{\'e}orie de la relativit{\'e} g{\'e}n{\'e}ralis{\'e}e (premi{\`e}re partie)(suite)},
  author={Cartan, {\'E}lie},
  booktitle={Annales scientifiques de l'{\'E}cole normale sup{\'e}rieure},
  volume={41},
  pages={1--25},
  year={1924}
}

@article{Charron2014,
  author  = {Charron, M. and Zadra, A. and Girard, C.},
  title   = {Four-dimensional tensor equations for a classical fluid in an external
             gravitational field},
  journal = {Q. J. R. Meteorol. Soc.},
  year    = {2014},
  volume  = {140},
  pages   = {908--916},
  url     = {https://doi.org/10.1002/qj.2185}
}

@article{Duval1985,
  author  = {Duval, C. and Burdet, G. and K{\"u}nzle, H. P. and Perrin, M.},
  title   = {Bargmann structures and {Newton--Cartan} theory},
  journal = {Phys. Rev. D},
  year    = {1985},
  volume  = {31},
  pages   = {1841--1853}
}

@article{Duval1991,
  author  = {Duval, C. and Gibbons, G. W. and Horv{\'a}thy, P.},
  title   = {Celestial mechanics, conformal structures and gravitational waves},
  journal = {Phys. Rev. D},
  year    = {1991},
  volume  = {43},
  pages   = {3907--3922}
}

@article{Eisenhart1928,
  author  = {Eisenhart, L. P.},
  title   = {Dynamical trajectories and geodesics},
  journal = {Ann. Math.},
  year    = {1928},
  volume  = {30},
  pages   = {591--606}
}

@book{hirsch2007numerical,
  title={Numerical computation of internal and external flows: The fundamentals of computational fluid dynamics},
  author={Hirsch, Charles},
  year={2007},
  publisher={Elsevier}
}

@book{isham1989modern,
  title={Modern differential geometry for physicists},
  author={Isham, Chris J},
  volume={32},
  year={1989},
  publisher={World Scientific}
}

@book{lamb1945hydrodynamics,
  title     = {Hydrodynamics},
  author    = {Lamb, Horace},
  year      = {1945},
  publisher = {Dover Publications},
  address   = {New York},
  pages     = {14},
  note      = {Reissue of the 6th edition (1932)},
  url       = {https://archive.org/details/hydrodynamics00lamb/page/14/mode/2up}
}

@article{Phillips1966,
  author  = {Phillips, N. A.},
  title   = {The equations of motion for a shallow rotating atmosphere
             and the `traditional approximation'},
  journal = {J. Atmos. Sci.},
  year    = {1966},
  volume  = {23},
  pages   = {626--628}
}

@techreport{pudykiewicz2017metric,
  author      = {Pudykiewicz, Janusz A.},
  title       = {Metric properties of the terrain following coordinates
                 on ellipsoid with mountains},
  institution = {Recherche en Pr\'{e}vision Num\'{e}rique Atmosph\'{e}rique},
  address     = {Environnement et Changement climatique Canada},
  year        = {2017},
  note        = {Available at \url{https://www.researchgate.net/publication/316669106}}
}

@incollection{ringler2011momentum,
  title={Momentum, vorticity and transport: Considerations in the design of a finite-volume dynamical core},
  author={Ringler, Todd D},
  booktitle={Numerical techniques for global atmospheric models},
  pages={143--183},
  year={2011},
  publisher={Springer}
}

@techreport{Vinokur1974,
  author      = {Vinokur, M.},
  title       = {A new formulation of the conservation equations of fluid dynamics},
  institution = {NASA Ames Research Center},
  year        = {1974},
  number      = {NASA-TM-X-62415},
  address     = {Moffett Field, CA}
}

@article{banerjee2016null,
  title={Null fluids: A new viewpoint of {Galilean} fluids},
  author={Banerjee, Nabamita and Dutta, Suvankar and Jain, Akash},
  journal={Physical Review D},
  volume={93},
  number={10},
  pages={105020},
  year={2016},
  publisher={APS}
}

@article{gal1975use,
  title   = {On the use of a coordinate transformation for the solution of the {Navier--Stokes} equations},
  author  = {Gal-Chen, Tzvi and Somerville, Richard C. J.},
  journal = {Journal of Computational Physics},
  volume  = {17},
  number  = {2},
  pages   = {209--228},
  year    = {1975},
  doi     = {10.1016/0021-9991(75)90037-6}
}

@article{laprise1992euler,
  title   = {The {Euler} equations of motion with hydrostatic pressure as an independent variable},
  author  = {Laprise, Ren{\'e}},
  journal = {Monthly Weather Review},
  volume  = {120},
  number  = {1},
  pages   = {197--207},
  year    = {1992},
  doi     = {10.1175/1520-0493(1992)120<0197:TEEOMW>2.0.CO;2}
}

@article{schar2002new,
  title   = {A new terrain-following vertical coordinate formulation for atmospheric prediction models},
  author  = {Sch{\"a}r, Christoph and Leuenberger, Daniel and Fuhrer, Oliver and L{\"u}thi, Daniel and Girard, Claude},
  journal = {Monthly Weather Review},
  volume  = {130},
  number  = {10},
  pages   = {2459--2480},
  year    = {2002}
}

@article{klemp2011terrain,
  title   = {A terrain-following coordinate with smoothed coordinate surfaces},
  author  = {Klemp, Joseph B.},
  journal = {Monthly Weather Review},
  volume  = {139},
  number  = {7},
  pages   = {2163--2169},
  year    = {2011},
  doi     = {10.1175/MWR-D-10-05046.1}
}

\appendix
\section{Integral Form of the Balance Laws}
\label{sec:gauss}

The equations derived in Section~\ref{sec:tensor} are already written in a form that is well suited to integral discretization methods. In particular, the expanded master equation
\begin{equation}
  \frac{1}{\Gdet}
  \frac{\partial}{\partial x^A}\left(\Gdet\,T^{AB}\right)
  + \Gamma^B_{AC}T^{AC}
  = 0
\end{equation}
separates naturally into a flux-divergence term and a geometric source term. The derivative with respect to $x^4$ vanishes because all physical fields are independent of $x^4$. Using $x^0=t$, we obtain
\begin{equation}
  \frac{\partial}{\partial t}\left(\Gdet\,T^{0B}\right)
  +
  \frac{\partial}{\partial x^i}\left(\Gdet\,T^{iB}\right)
  =
  -\Gdet\,\Gamma^B_{AC}T^{AC},
  \qquad
  B=0,1,2,3,4.
  \label{eq:master_flux_source}
\end{equation}

This form has the structure familiar from finite-volume, discontinuous Galerkin, and spectral element methods: the first term describes the local change of the quantity represented by $T^{0B}$, the second describes its spatial flux, and the right-hand side contains the geometric source associated with the connection.

Let $V$ denote a control volume fixed in the chosen spatial coordinates, with boundary $\partial V$. Its physical image may move under a time-dependent coordinate mapping. Integrating \eqref{eq:master_flux_source} over $V$ gives
\begin{equation}
  \frac{\mathrm d}{\mathrm dt}
  \int_V
  \Gdet\,T^{0B}\,\mathrm d^3x
  +
  \int_V
  \frac{\partial}{\partial x^i}\left(\Gdet\,T^{iB}\right)\mathrm d^3x
  =
  -\int_V
  \Gdet\,\Gamma^B_{AC}T^{AC}\,\mathrm d^3x.
  \label{eq:master_volume}
\end{equation}

The spatial divergence in the second term can be converted into a flux through the boundary of the control volume using the Gauss divergence theorem. Let $n_i$ denote the ordinary outward unit normal and $\mathrm dS_{\rm coord}$ the ordinary coordinate-space surface element on $\partial V$, so that the divergence theorem applied to the densitized flux $\Gdet\,T^{iB}$ reads
\begin{equation}
  \int_V
  \frac{\partial}{\partial x^i}\left(\Gdet\,T^{iB}\right)\mathrm d^3x
  =
  \oint_{\partial V}
  \Gdet\,T^{iB}\,n_i\,\mathrm dS_{\rm coord}.
\end{equation}
It is convenient to absorb the volume factor into the surface element itself by defining the densitized surface element
\begin{equation}
  \mathrm d\Sigma_i \equiv \Gdet\,n_i\,\mathrm dS_{\rm coord},
\end{equation}
so that
\begin{equation}
  \int_V
  \frac{\partial}{\partial x^i}\left(\Gdet\,T^{iB}\right)\mathrm d^3x
  =
  \oint_{\partial V}
  T^{iB}\,\mathrm d\Sigma_i.
  \label{eq:gauss_tensor_flux}
\end{equation}
Combining the volume and boundary terms gives the integral form of the master equation:
\begin{equation}
  \frac{\mathrm d}{\mathrm dt}
  \int_V
  \Gdet\,T^{0B}\,\mathrm d^3x
  +
  \oint_{\partial V}
  T^{iB}\,\mathrm d\Sigma_i
  =
  -\int_V
  \Gdet\,\Gamma^B_{AC}T^{AC}\,\mathrm d^3x
  \label{eq:master_integral}
\end{equation}
for each $B=0,1,2,3,4$.

Equation~\eqref{eq:master_integral} is often the starting point for numerical discretization, in which the boundary integral can be replaced by a sum of numerical fluxes across the cell faces, evaluated against $\mathrm d\Sigma_i$, while the source term is approximated by an appropriate volume quadrature.

\section{List of Symbols}
\label{sec:symbols}

\begingroup
\small
\renewcommand{\arraystretch}{1.15}
\begin{longtable}{@{}p{0.24\textwidth}p{0.71\textwidth}@{}}
\toprule
Symbol & Description \\
\midrule
\endfirsthead
\toprule
Symbol & Description (continued) \\
\midrule
\endhead
\bottomrule
\endfoot
\multicolumn{2}{@{}l}{\textbf{Coordinates, indices, and conventions}}\\
$A,B,C,\ldots$ & Five-dimensional indices, ranging over $0,1,2,3,4$.\\
$\mu,\nu,\ldots$ & Spacetime indices, ranging over $0,1,2,3$.\\
$i,j,k,\ldots$ & Spatial indices, ranging over $1,2,3$.\\
$x^A$, $x^A(t)$ & Coordinates and the coordinates of a parcel trajectory; $x^0=t$ and $x^4$ is auxiliary.\\
$t$ & Absolute Newtonian time.\\
$\lambda,\varphi,r$ & Rotating longitude, latitude, and distance from the planetary centre.\\
$\eta$ & Generalized vertical coordinate, with mapping $r=r(t,\lambda,\varphi,\eta)$.\\
$\dot\eta$ & Vertical coordinate velocity, $\mathrm d\eta/\mathrm dt$, given by \eqref{eq:eta_velocity}.\\
$r_t,r_\lambda,r_\varphi,r_\eta$ & Partial derivatives of the vertical mapping in the new coordinates.\\
$a,z$ & Reference radius and elevation $z=r-a$.\\
$r_s,r_T$ & Terrain radius and model-top radius in the example vertical mapping.\\
$\tilde{\phantom{x}}$, $\hat{\phantom{x}}$ & Transformed quantities; hats denote generalized vertical coordinates.\\
$(a)$ & Superscript identifying the approximate thin-shell metric and potential.\\
\addlinespace
\multicolumn{2}{@{}l}{\textbf{Geometry and kinematics}}\\
$\mathcal M$ & Smooth manifold.\\
$g_{AB},g^{AB}$ & Five-dimensional metric and its inverse; $g^{ij}$ is also the inverse of $g_{ij}$.\\
$\delta_{ij},\delta^{ij}$ & Euclidean spatial metric components and their inverse in Cartesian coordinates.\\
$\delta^A{}_B$ & Kronecker delta.\\
$\Gdet$ & Metric volume factor; here equal to $\sqrt{\det(g_{ij})}$.\\
$\mathcal J$ & Volume factor in generalized vertical coordinates, $|\partial r/\partial\eta|r^2\cos\varphi$.\\
$\Gamma^A_{BC},\nabla_A$ & Levi--Civita connection coefficients and covariant derivative.\\
$\mathrm D/\mathrm Dt$ & Covariant derivative along a parcel trajectory; for scalars, the material derivative.\\
$\mathrm d/\mathrm dt$ & Ordinary derivative of components and coordinates along a parcel trajectory.\\
$u^A,u_A$ & Five-velocity components and their metric-lowered form; $u^0=u_4=1$.\\
$u^i$ & Spatial coordinate velocities, $\mathrm dx^i/\mathrm dt$.\\
$u^4$ & Auxiliary velocity component, fixed by the null condition: $u^4=-L$.\\
$u_0$ & Covariant time component, $u_0=-E_{\mathrm{mech}}$.\\
$v_i$ & Coordinate-relative covariant velocity, $v_i=g_{ij}u^j=u_i-g_{0i}$.\\
$\bar u^A$ & Reference motion orthogonal to spatial coordinate vectors; $\bar u^i=-g^{ij}g_{0j}$.\\
$\xi^A$ & Covariantly constant vector tangent to the auxiliary direction, $\xi^A=\delta^A_4$.\\
$\mathcal P^{AB}$ & Pressure projector relative to the fluid velocity, $g^{AB}-u^A\xi^B-\xi^Au^B$.\\
$\alpha_i$ & Prescribed covector components subtracted from $u_i$ in generalized relative forms.\\
\addlinespace
\multicolumn{2}{@{}l}{\textbf{Fluid and mechanical quantities}}\\
$\rho,p,e$ & Mass density, pressure, and specific internal energy.\\
$T^{AB}$ & Symmetric mass--momentum tensor including internal energy and pressure work.\\
$L,K$ & Lagrangian and kinetic energy per unit mass.\\
$E_{\mathrm{mech}},E$ & Specific mechanical energy $K+\Phie$ and total energy $e+E_{\mathrm{mech}}$.\\
$\Phi$ & Scalar Newtonian gravitational potential.\\
$\Phie$ & Effective potential, $\Phi-\tfrac12g^{ij}g_{0i}g_{0j}=-g_{00}/2$.\\
$\hat\Phie$ & Effective potential defined by the transformed metric, $-\hat g_{00}/2$.\\
$U,V,W$ & Eastward, northward, and radial or vertical speeds relative to the rotating frame.\\
$w$ & Physical radial velocity in generalized vertical coordinates, $\mathrm dr/\mathrm dt$; distinct from $\mathrm d\eta/\mathrm dt$.\\
\addlinespace
\multicolumn{2}{@{}l}{\textbf{Atmospheric and planetary quantities}}\\
$T,\theta,\pi$ & Temperature, potential temperature, and Exner function; $T=\theta\pi$.\\
$p_0$ & Reference pressure defining the Exner function.\\
$R_d,c_{pd},c_{vd}$ & Dry-air gas constant and specific heat capacities at constant pressure and volume.\\
$\gamma_d$ & Ratio of heat capacities, $c_{pd}/c_{vd}$.\\
$h,s$ & Specific enthalpy and entropy.\\
$G,M$ & Newtonian gravitational constant and planetary mass.\\
$\Omega,f$ & Planetary angular speed and traditional Coriolis parameter $f=2\Omega\sin\varphi$.\\
$g_r$ & Constant effective gravity; $r$ is a label, not a tensor index or summation index.\\
\addlinespace
\multicolumn{2}{@{}l}{\textbf{Integral balances}}\\
$V,\partial V$ & Control volume fixed in the chosen spatial coordinates and its boundary.\\
$n_i,\mathrm dS_{\rm coord}$ & Outward unit normal and surface element in coordinate space.\\
$\mathrm d\Sigma_i$ & Densitized surface element, $\Gdet\,n_i\,\mathrm dS_{\rm coord}$.\\
\end{longtable}
\endgroup

\end{document}